\documentclass{aastex63}

\usepackage{amsmath}
\usepackage{tablefootnote}
\usepackage[utf8]{inputenc}
\usepackage[default,regular,black]{sourceserifpro}
\usepackage[T1]{fontenc}

\usepackage{txfonts}

\newcommand{\mearth}{\,M$_\oplus$}

\newcommand\ah{}

\usepackage{xcolor}

\graphicspath{{./}{figures_submission/}}

\accepted{on October 21 2022}

\shorttitle{Sample article}
\shortauthors{Hales et al.}

\begin{document}

\title{ALMA Observations of the HD~110058  debris disk }

\author{Antonio S. Hales}
\affiliation{Joint ALMA Observatory, Avenida Alonso de C\'ordova 3107, Vitacura 7630355, Santiago, Chile}
\affiliation{National Radio Astronomy Observatory, 520 Edgemont Road, Charlottesville, VA 22903-2475, United States of America}
\author{Sebasti\'an Marino}
\affiliation{Jesus College, University of Cambridge, Jesus Lane, Cambridge CB5 8BL, UK}
\affiliation{Institute of Astronomy, University of Cambridge, Madingley Road, Cambridge CB3 0HA, UK}
\author{Patrick D. Sheehan}
\affiliation{Center for Interdisciplinary Exploration and Research in Astronomy,
  Northwestern University, 1800 Sherman Ave.}
\affiliation{National Radio Astronomy Observatory, 520 Edgemont Road, Charlottesville, VA 22903-2475, United States of America}
\author{Silvio Ulloa}
\affiliation{Departamento de Astronom\'ia, Universidad de Chile, Casilla 36-D, Santiago 8330015, Chile}
\author{Sebasti\'an P\'erez}
\affiliation{Departamento de F\'isica, Universidad de Santiago de Chile, Av. Ecuador 3493, Estaci\'on Central, Santiago, Chile}
\affiliation{Millennium Nucleus on Young Exoplanets and their Moons (YEMS), Chile}
\affiliation{Center for Interdisciplinary Research in Astrophysics and Space Exploration (CIRAS), Universidad de Santiago de Chile, Santiago, Chile}
\author{Luca Matr\`a}
\affiliation{School of Physics, Trinity College Dublin, the University of Dublin, College Green, Dublin 2, Ireland}
\affiliation{School of Physics, National University of Ireland Galway, University Road, Galway H91 TK33, Ireland}
\author{Quentin Kral}
\affiliation{LESIA, Observatoire de Paris, Universit\'e PSL, CNRS, Sorbonne Universit\'e, Univ. Paris Diderot, Sorbonne Paris Cit\'e, 5 place Jules Janssen, 92195 Meudon, France}
\author{Mark Wyatt}
\affiliation{Institute of Astronomy, University of Cambridge, Madingley Road, Cambridge CB3 0HA, UK}
\author{William Dent}
\affiliation{Joint ALMA Observatory, Avenida Alonso de C\'ordova 3107, Vitacura 7630355, Santiago, Chile}
\affiliation{European Southern Observatory, Avenida Alonso de C\'{o}rdova 3107, Vitacura 7630355, Santiago, Chile}
\author{John Carpenter}
\affiliation{Joint ALMA Observatory, Avenida Alonso de C\'ordova 3107, Vitacura 7630355, Santiago, Chile}
\affiliation{National Radio Astronomy Observatory, 520 Edgemont Road, Charlottesville, VA 22903-2475, United States of America}

%
%

%
%
%
\begin{abstract}

We present Atacama Large Millimeter Array (ALMA) observations of the
young, gas-rich debris disk around HD110058 at 0.3-0.6\arcsec
resolution.  The disk is detected in the 0.85 and 1.3~mm continuum, as
well as in the J=2-1 and J=3-2 transitions of $^{12}$CO and
$^{13}$CO. The observations resolve the dust and gas distributions and
reveal that this is the smallest debris disk around stars of similar
luminosity observed by ALMA. The new ALMA data confirm the disk is
very close to edge-on, as shown previously in scattered light images.
We use radiative transfer modeling to constrain the physical
properties of dust and gas disks. The dust density peaks at around
31~au and has a smooth outer edge that extends out to
$\sim70$~au. Interestingly, the dust emission is marginally resolved
along the minor axis, which indicates that it is vertically thick if
truly close to edge-on with an aspect ratio between 0.13 and 0.28. We
also find that the CO gas distribution is more compact than the dust
\ah{(similarly to the disk around 49 Ceti)}, which could be due to a low
viscosity and a higher gas release rate at small radii. Using
simulations of the gas evolution taking into account the CO
photodissociation, shielding, and viscous evolution, we find that
HD~110058's CO gas mass and distribution are consistent with a
secondary origin scenario. Finally, we find that the gas densities may
be high enough to cause the outward drift of small dust grains in the
disk.

\end{abstract}






\section{Introduction}\label{intro}

The formation and evolution of planetary systems is a central question
of modern astrophysics. During the past decade, the Atacama Large
Millimeter/submillimeter Array (ALMA) has revolutionized our
understanding of the star and planet \ah{formation} process. ALMA has
pierced into all-star/planet formation stages. It has discovered the
earliest stages of disk formation, exposed rich details of
planet-forming disks, and unveiled the architecture and dynamics of
the aftermath of planetary formation: the debris disk stage.

Debris disks are main-sequence stars surrounded by
dust. High-resolution observations at different wavelengths show that
debris disks are usually composed of one or more rings of dust
\citep{hughes2018}.  The dust rings are not remnants of the planet
formation process but instead created by collisions between
comet-sized or larger bodies \citep{wyatt2008,wyatt2015}.  The
dust-to-star luminosity ratio or fractional luminosity is a measure of
the amount of dust.

ALMA millimeter and sub-millimeter observations show  that debris disks with high fractional luminosities and ages between 10-40~Myr show
high levels of cold CO gas \citep{moor2017}.
The origin and evolution of the gas in debris disk is unclear
\citep{hughes2018,moor2020}. The main question about the origin is
whether the gas is primordial (i.e., leftover from the protoplanetary
disk phase) or produced by in-situ by collisions (i.e.,
second-generation or secondary). Some disks seem to be massive enough
to shield the CO from stellar and interstellar UV-photodissociation
\citep{Kospal13,pericaud2017}. These disks support the primordial
origin scenario.  Other disks (like $\beta$~Pic), have too low amounts
of gas to shield CO, and therefore, the gas observed must be released
continuously by collisions of ice-rich bodies in the cold planetesimal
belts \citep[e.g.,][]{kral2016,marino2016,matra2017a,matra2017b}.
As CO is released from solid bodies, the photodissociation of CO will
produce neutral carbon which in turn can prevent the remaining CO from
being photodissociated \citep{kral2019}. This mechanism can also explain the presence of large amount of gas in young systems where the primordial scenario is not the viable explanation \citep{hales2019,cataldi2020}, and provides
predictions for the physical and chemical evolution of the gas
\citep{Marino2020}. The evolutionary pathways each disk will follow
are highly dependent of the properties of each system, such as stellar
UV, gas release rate and disk viscosity \citep{Marino2020}.

These models can also provide important clues on the volatile
composition of exocomets in the outer regions of planetary systems, in
systems undergoing the dynamically active final stages of terrestrial
planet formation when volatile delivery events are most likely to
happen \citep[e.g.][]{Rubin2019}. Further studies of additional
gas-rich debris disks are needed to establish the general case on the
origin of the gas.

HD~110058 is a 17~Myr old A0V star that harbors a debris disk
\citep{Mannings1998}.  It is part of the Lower-Centaurus-Crux (LCC) association
and is located at a distance of  129.9$_{-1.2}^{+1.3}$~pc \citep{gaia2018,goldman2018}.

The disk is close to edge-on and detected in scattered light in virtue
of its high surface brightness \citep{kasper2015, esposito2020}. The high
inclination also enables the detection of atomic gas observed in
absorption \citep{hales2017, Rebollido2018}. The 1.3mm continuum and
$^{12}$CO(2-1) luminosities reported in previous ALMA data are also
similar to those of $\beta$~Pic \citep{Lieman2016,moor2020}.
HD~110058 is the only known young A-star with similar gas and dust
emission levels and high inclination as in $\beta$ Pictoris. For instance, the
$^{12}$CO(2-1) emission of HD~110058 is three times more luminous than
that of HD~181327 and 30\% more luminous than that of Fomalhaut (after correcting their fluxes by $1/d^2$); both
these debris disks have gas and have evidence of a cometary origin for
their gas content.

This work presents new ALMA observations at 0.3-0.6\arcsec resolution of the
debris disk around HD~110058.  The band 6 and 7 data, at 1.32~mm and
0.88~mm, respectively, are presented in Section~\ref{obs}, including
the description of the continuum and CO gas observations. We describe
the imaging results in Section \ref{results}. The modeling we
use to fit the continuum, and the spectral line kinematics is shown
in Section~\ref{rtmodel}. A discussion conveying our interpretation
of the gas and dust observations of HD 110058 is presented in
Section~\ref{discussion}.

\section{Observations and Data Reduction}\label{obs}

\begin{deluxetable*}{lccccccccc}
\tablecaption{Summary of ALMA Observations \label{log1}}
\tablewidth{700pt}
\tabletypesize{\scriptsize}
\tablehead{
\colhead{Band} & 
\colhead{Execution Block} & 
\colhead{N Ant.} & 
\colhead{Date} & 
\colhead{ToS } & 
\colhead{Avg. Elev. } & 
\colhead{Mean PWV } & 
\colhead{Baseline } & 
\colhead{AR } & 
\colhead{MRS} \\
\colhead{} & 
\colhead{} & 
\colhead{} & 
\colhead{} & 
\colhead{(sec)} & 
\colhead{(deg)} & 
\colhead{(mm)} & 
\colhead{(m)} & 
\colhead{(\arcsec)} & 
\colhead{(\arcsec)} 
}
\startdata
Band 6 &uid://A002/Xdb6217/X4488&46&2019-04-27&4024&63.3&1.2  &15-740  &0.4 &5.6 \\
Band 6 &uid://A002/Xdb6217/X4b0b&46&2019-04-27&4023&58.4 &0.8 &15-740 &0.4 &5.6 \\
Band 6 &uid://A002/Xdb7ab7/X1373&46&2019-04-28&3994&47.3 &0.8 &15-783 &0.4 &5.7\\
Band 7 &uid://A002/Xdb6217/X55ec&46&2019-04-27&4676&42.3&0.7  &15-740&0.3&3.8\\
Band 7 &uid://A002/Xdb7ab7/X58c &46&2019-04-28&4859&59.5&0.8  &15-783&0.3&3.6\\
Band 7 &uid://A002/Xdb7ab7/Xa3a &49&2019-04-28&4663&63.4&0.8  &15-783&0.3&3.6\\
Band 7 &uid://A002/Xdb7ab7/Xd39 &46&2019-04-28&4642&56.6&0.8  &15-783&0.3&3.8\\
\enddata

\tablecomments{Summary of the new ALMA observations presented in this
  work. The table shows the total number of antennas, total time on
  source (ToS), target average elevation, mean precipitable water
  vapor column (PWV) in the atmosphere, minimum and maximum baseline
  lengths, expected angular resolution (AR) and maximum recoverable
  scale (MRS).}
\end{deluxetable*}


ALMA observations of HD~110058 were acquired in the nights of April
27$^{th}$ and 28$^{th}$ 2019 using the Band~6 and Band~7 receivers
(Project code 2018.1.00500.S). The total number of available
12 meter antennas ranged from 46 to 49, providing baselines between
15.1~m to 783~m.  A summary of the observations is presented in
Table~\ref{log1}. Standard observations of bandpass, flux and phase
calibrators were also included.

The correlator setup for Band 6 observations included two Frequency
Division Mode (FDM) spectral windows tuned to cover the $^{12}$CO and
$^{13}$CO ($J = 2-1$) transitions with a spectral resolution of
0.564~MHz ($\sim$0.75 ~km~s$^{-1}$) and 937.5~MHz bandwidth.  Two
Time Division Mode (TDM) spectral windows were dedicated to continuum
measurements, each providing a total bandwidth of $1.875$ GHz. The
Band 7 observations used a similar strategy. Two FDM spectral windows
were tuned to cover the $^{12}$CO and $^{13}$CO ($J = 3-2$)
transitions.  The $^{12}$CO(3-2) spectral window had spectral
resolution of 0.564~MHz ($\sim$0.49 ~km~s$^{-1}$) and 937.5~MHz
bandwidth, while the $^{13}$CO(3-2) spectral window had spectral
resolution of 0.468~MHz ($\sim$0.26 ~km~s$^{-1}$) and 468.75~MHz
bandwidth.
 
%

The data was calibrated using the ALMA Science Pipeline (version
42254M Pipeline-CASA54-P1-B) in CASA 5.4.0
\citep[CASA{\footnote{\url{http://casa.nrao.edu/}}};][]{McMullin2007}
by ALMA staff. The calibration process includes correction from Water
Vapor Radiometer (WVR) data, system temperature, as well as bandpass,
phase, and amplitude calibrations of the interferometric data.

\begin{figure}[h!]
    \centering
    \includegraphics[scale=0.32]{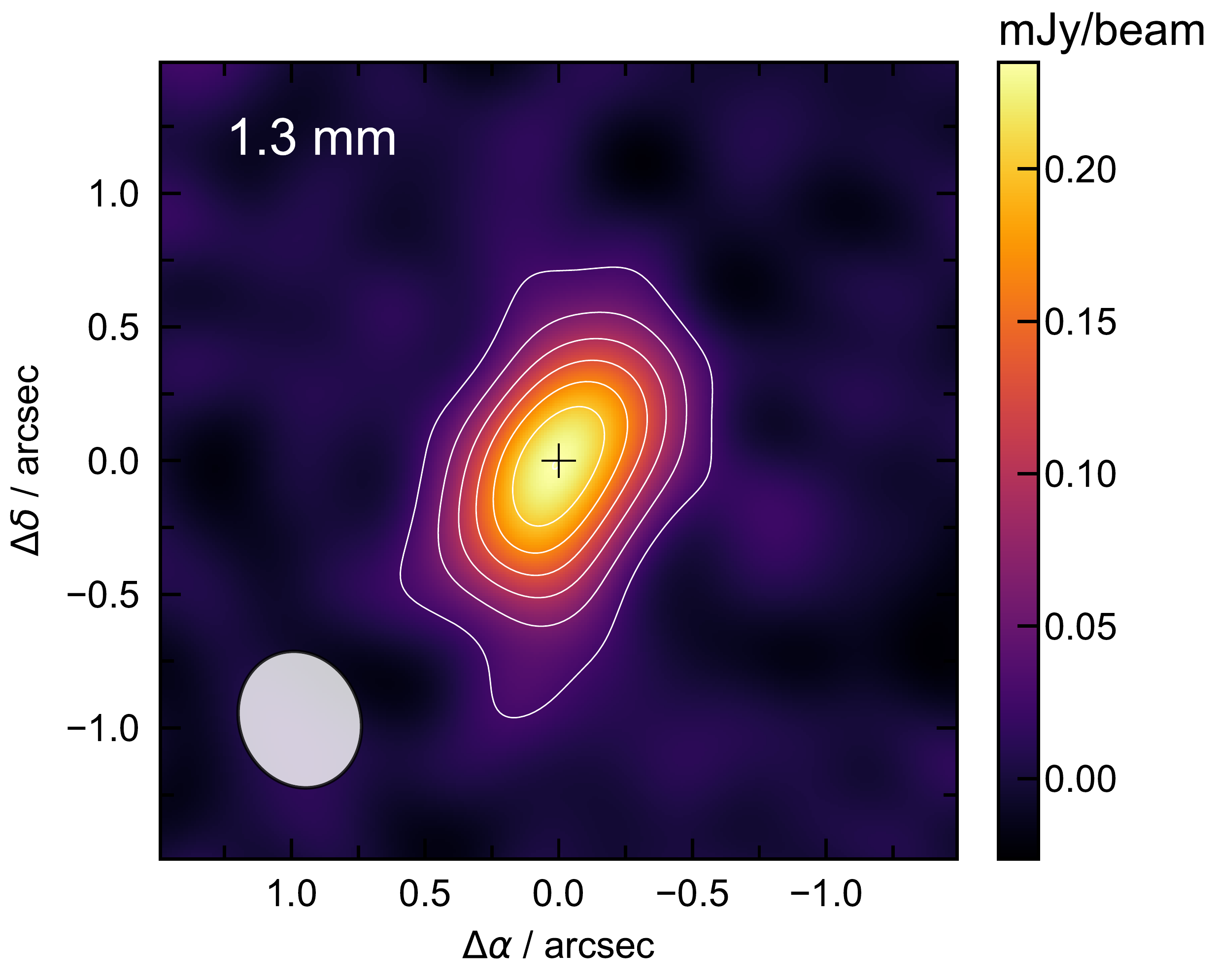}
    \includegraphics[scale=0.32]{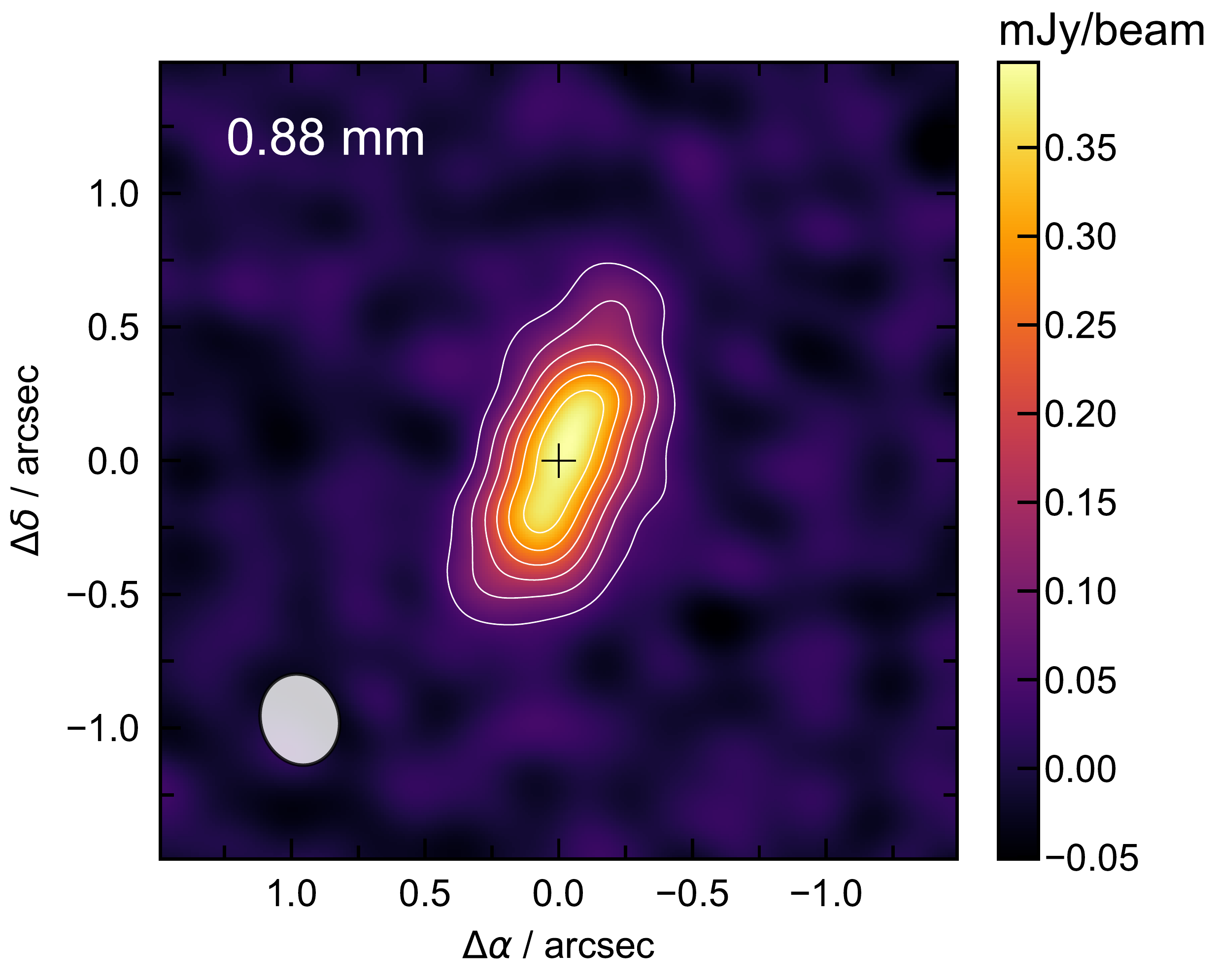}
    \caption{{ALMA 1.32 and 0.88~mm continuum images of the HD110058 disk (left and right panels). Contour levels are shown at 3, 7, 14 ,21 times the RMS of each image (8~ $\mu$Jy~beam$^{-1}$ and 17~$\mu$Jy~beam$^{-1}$ respectively). The position of the star is indicated with a black plus symbol. The integrated fluxes reported in Table~\ref{table-1} were integrated within a circular region centered at the star position of $1.0\arcsec$ in radius. The ellipse in the lower left represents the synthetic beam of the data. 
      }}
        \label{cont_images}
\end{figure}

Imaging of the continuum was performed using the {\sc TCLEAN} task in
CASA. The two continuum spectral window and the line-free channels of
the $^{12}$CO and $^{13}$CO spectral window were imaged together to
produce a single continuum image in each band.  An image centered at
225.75~GHz (1.328~mm) with total aggregate bandwidth of 5.61~GHz was
produced using Briggs weighting with a robust parameter of 0.5, which
yielded a synthesized beam size of 0.52$\arcsec$ $\times$
0.45$\arcsec$ at position angle (PA) of 68.1$^{\circ}$ (Figure
\ref{cont_images}). Similarly to Band 6, in the Band 7 data all
line-free channels were used in {\sc TCLEAN} to produce an image
centered at 338.30~GHz (0.886~mm) with total aggregate band-with of
4.99~GHz (see Fig. \ref{cont_images}). Using Briggs weighting with a
robust parameter of 0.5 resulted in a synthesized beam size of
0.34$\arcsec$ $\times$ 0.29$\arcsec$ (PA = 73.9$^{\circ}$). The
properties of the 1.32 and 0.88~mm continuum images are summarized in
Table~\ref{table-1}.

\begin{figure}[h!]
    \centering
    \includegraphics[scale=0.24]{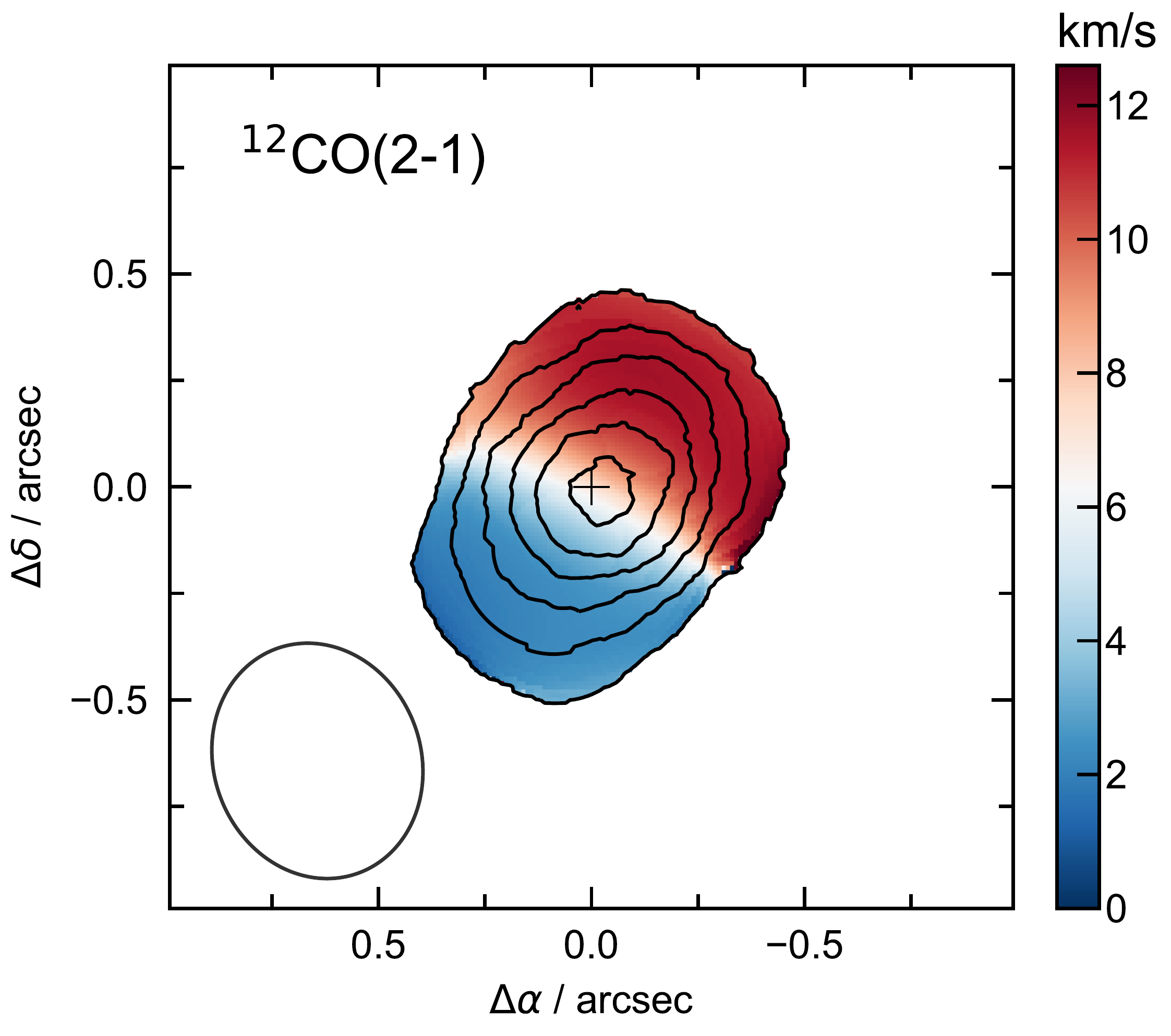}
    \includegraphics[scale=0.24]{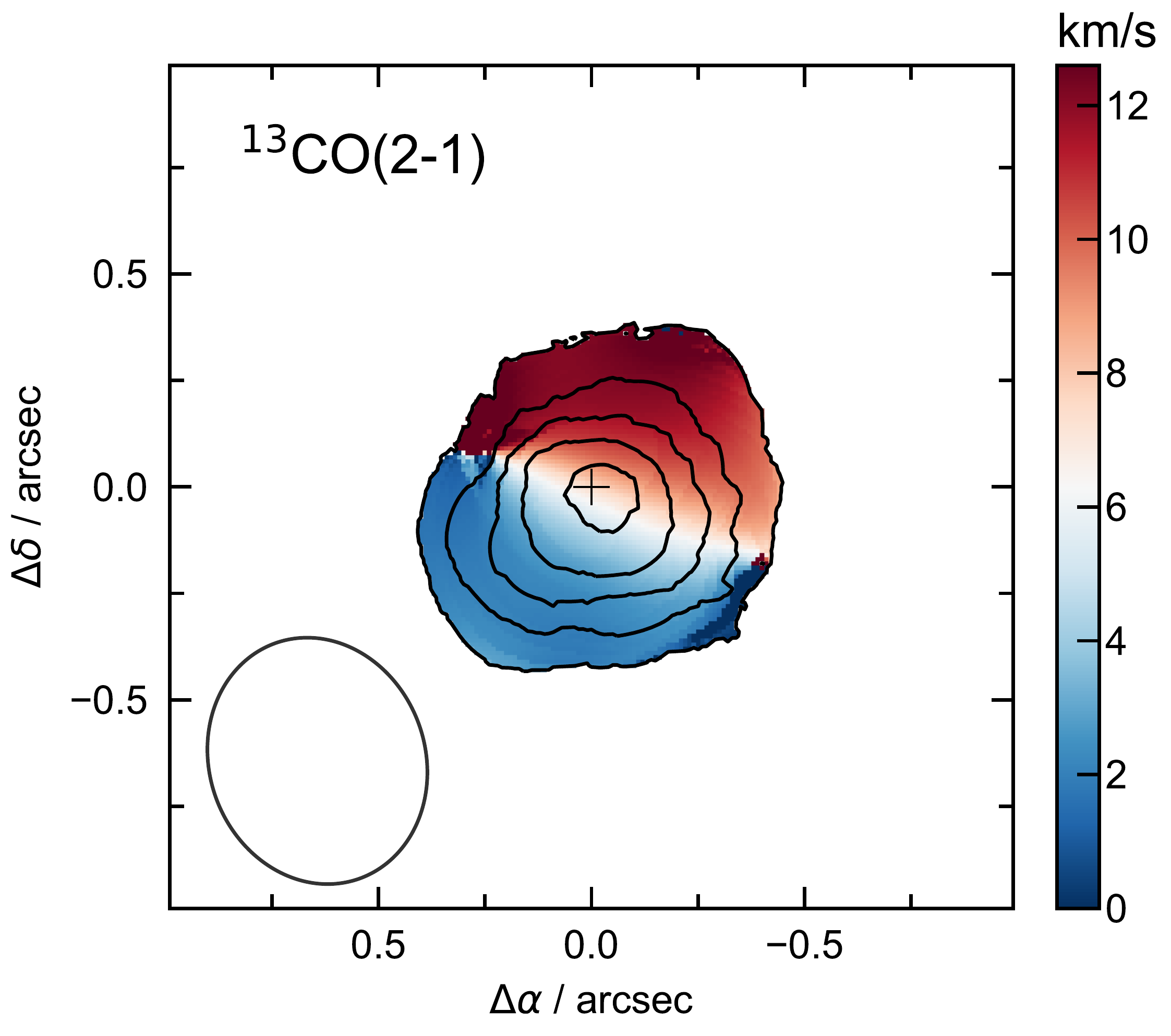}
     \includegraphics[width=9cm,angle=00,scale=0.715]{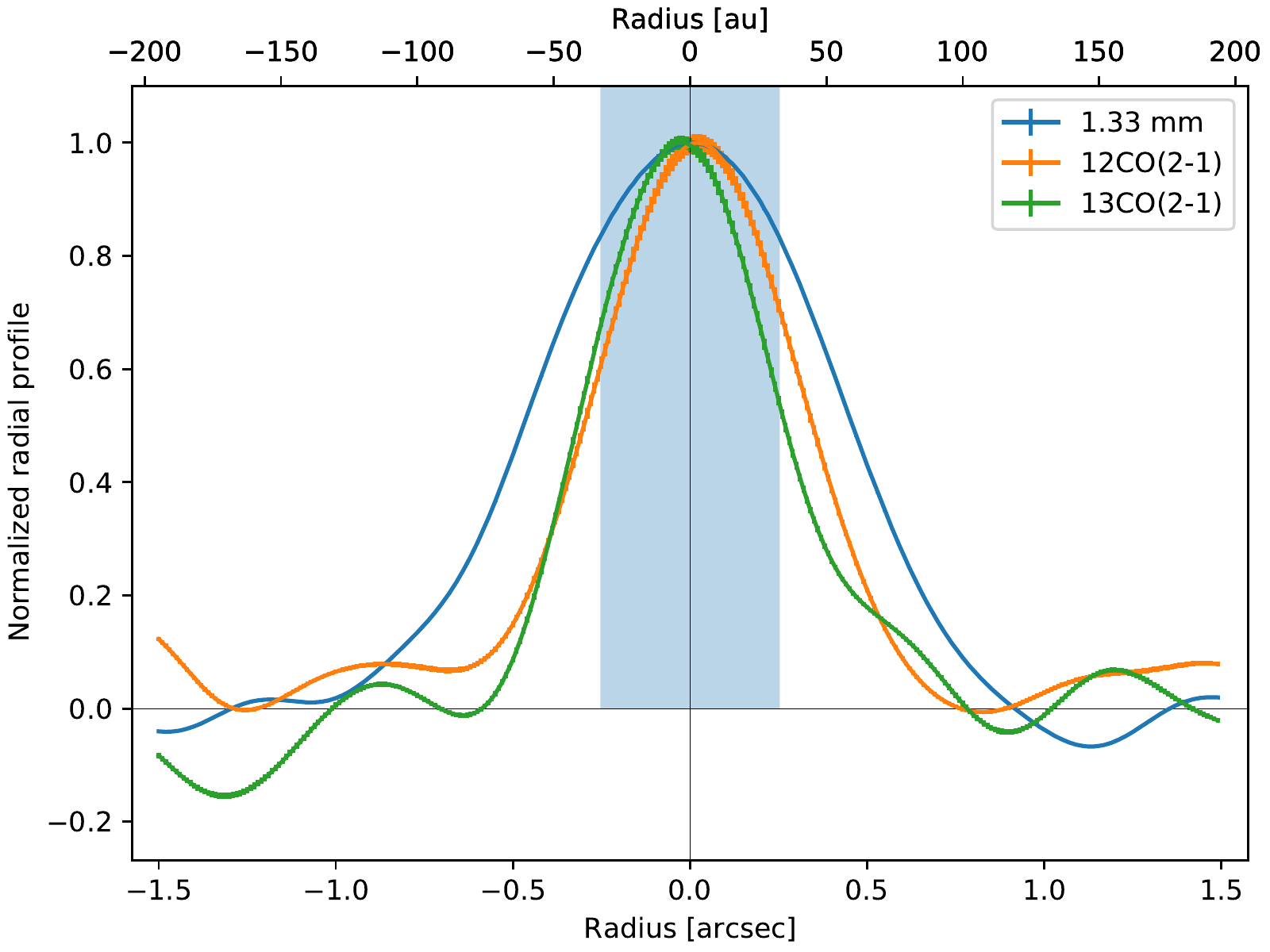}
        \caption{ $^{12}$CO (2$-$1) and $^{13}$CO(2$-$1) moment 0 and
          moment 1 maps of HD110058 (left and center panels,
          respectively). Colored maps show the velocity centroids for
          each pixel, i.e. moment~1.  Contours show integrated
          intensity of line emission. First contour is 3 times the RMS
          noise and each subsequent contour level increases with a
          step of 3$\times$RMS (i.e. contour levels are 3, 6, 9, 12,
          and so on, times the RMS).  The corresponding RMS noise for
          the zeroth moment are listed in Table~\ref{table-1}.
          {\bf{Right:}} Radial distribution of 1.33~mm continuum,
          $^{12}$CO (2$-$1) and $^{13}$CO (2$-$1) obtained from
          spatially integrating the continuum and moment 0 images
          along a line perpendicular to the major axis of the
          disk. The profiles have been normalised to unity for
          comparison.  The blue region shows the projection of the
          synthesized beam size in the direction of the disk's major
          axis.  }
        \label{band6-co}
\end{figure}

\begin{figure}[h!]
    \centering
    \includegraphics[scale=0.24]{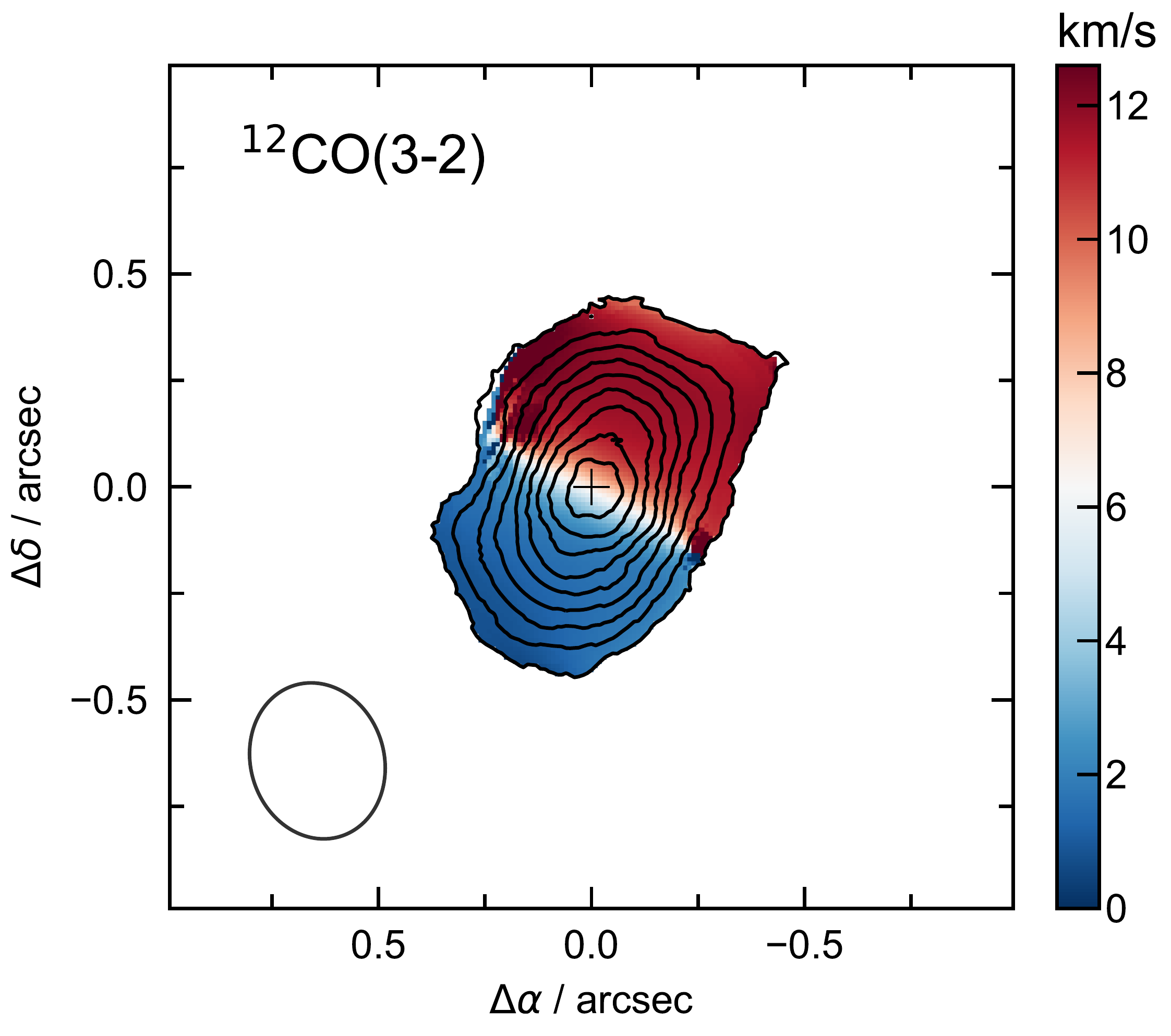}
    \includegraphics[scale=0.24]{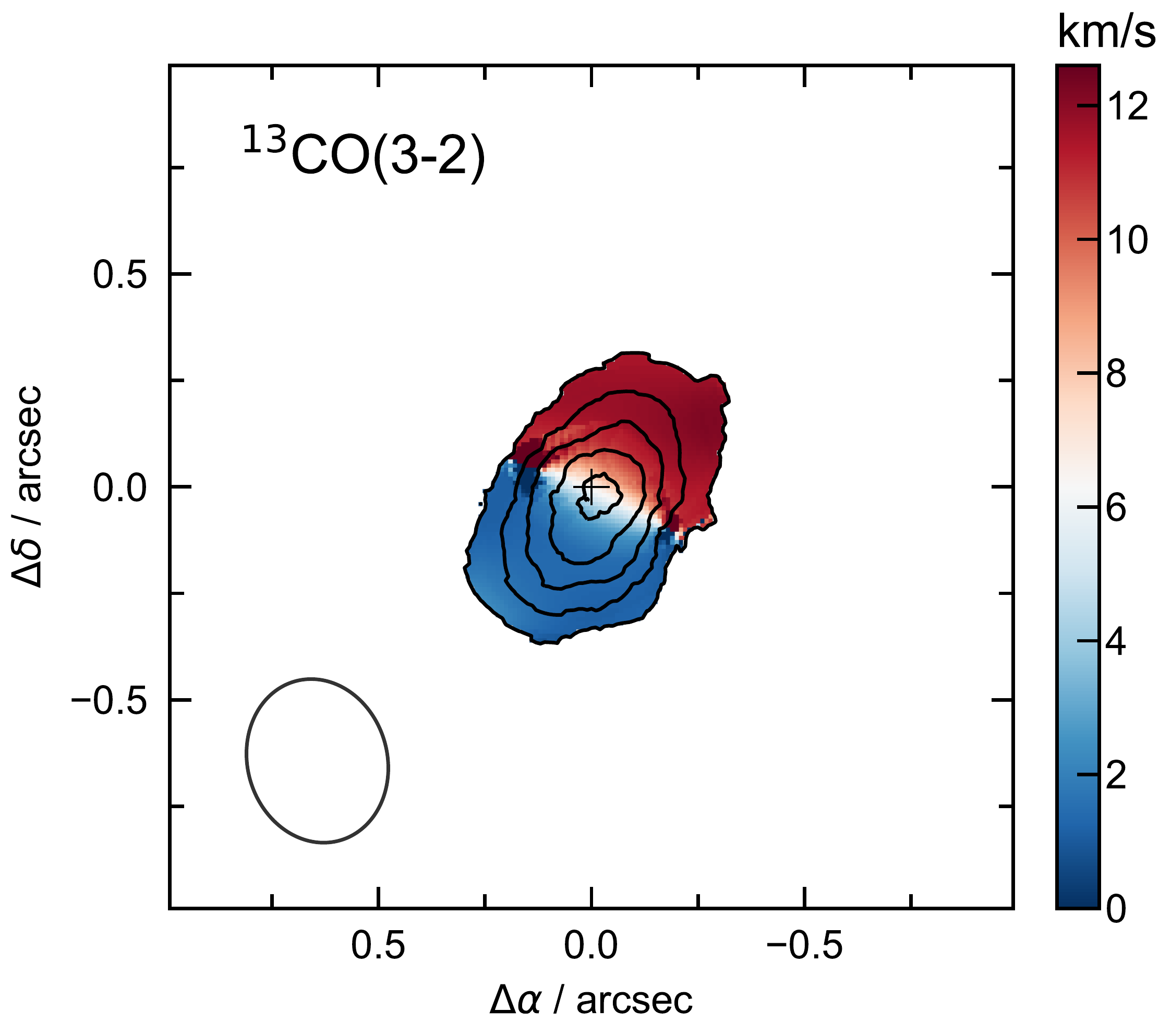}
     \includegraphics[width=9cm,angle=00,scale=0.715]{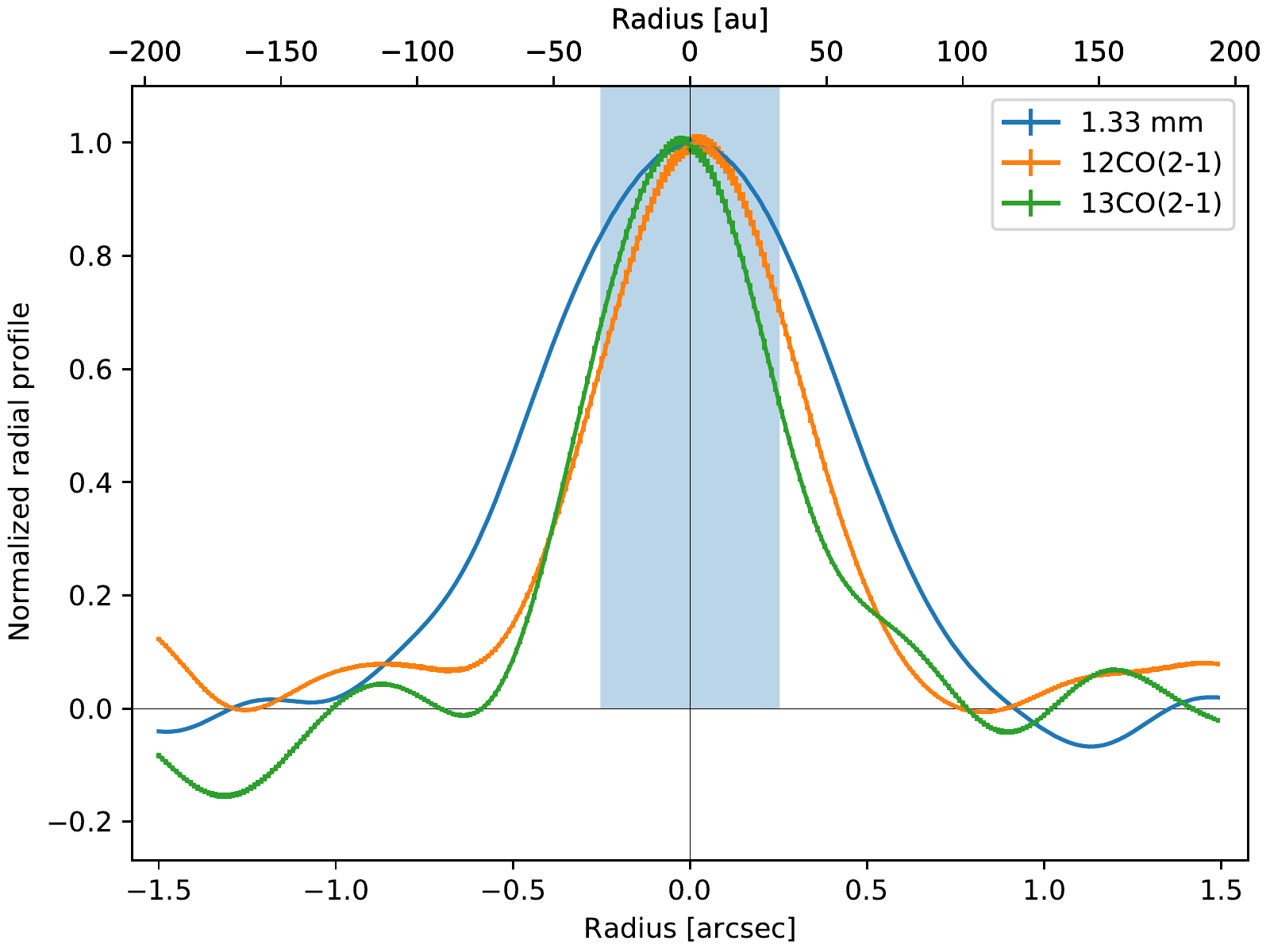}

        \caption{ $^{12}$CO(3$-$2) and $^{13}$CO(3$-$2) moment 0 and
          moment 1 maps of HD110058 (left and center panels,
          respectively).Colored maps show the velocity centroids for
          each pixel, i.e. moment~1.  Contours show integrated
          intensity of line emission. First contour is 3 times the RMS
          noise and each subsequent contour level increases with a
          step of 3$\times$RMS (i.e. contour levels are 3, 6, 9, 12,
          and so on, times the RMS).  The corresponding RMS noise for
          the zeroth moment of $^{12}$CO(3$-$2) and $^{13}$CO(3$-$2)
          are 3.4, and 4.9 mJy beam$^{-1}$ km~s$^{-1}$, respectively.
          {\bf{Right:}} Radial distribution of 0.88~mm continuum,
          $^{12}$CO (3$-$2) and $^{13}$CO (3$-$2) obtained from
          spatially integrating the continuum and moment 0 images
          along a line perpendicular to the major axis of the
          disk. The profiles have been normalised to unity for
          comparison. The blue region shows the projection of the
          synthesized beam size in the direction of the disk's  major axis. }
        \label{band7-co}
\end{figure}

Continuum subtraction in the visibility domain was performed prior to
imaging each molecular line, by fitting the continuum in the line-free
channels and subtracting it in the {\sc{uv}}-space using the task
UVCONTSUB. {\sc TCLEAN}ing of the line data was done using similar
parameters as for the continuum images. The spectral resolution of the
final cubes were 0.7~km~s$^{-1}$ for $^{12}$CO(2$-$1), 0.8~km~s$^{-1}$
for $^{13}$CO(2$-$1), 0.5 km s$^{-1}$ for $^{12}$CO(3$-$2) and 0.3~km
s$^{-1}$ for $^{13}$CO(3$-$2). The image properties for each data cube
are listed in Table~\ref{table-1}. Integrated intensity (moment 0)
maps for the $^{12}$CO and $^{13}$CO lines were produced with CASA
task {\sc IMMOMENTS}, and are shown in Figures~\ref{band6-co} and
\ref{band7-co}.  The moment 0 images were produced by integrating the
signal in channels with emission above 3$\sigma$, corresponding to
velocities between -5.0 and +17.5~km~s$^{-1}$ from the stellar
velocity (v$_{sys}\sim 6.4$ km~s$^{-1}$) for $^{12}$CO, and between
-3.5 and 15.2~km~s$^{-1}$ for $^{13}$CO.


\section{Results}\label{results}

\begin{table}
  \caption{Integrated Fluxes and Gaussian Fit Parameters}
  \label{table-1}
  \begin{center}
    \leavevmode
    \begin{tabular}{lcccccc} \hline \hline  
        Source Properties                           & Continuum Band 6 & Continuum Band 7   &  $^{12}$CO (2$-$1)        &  $^{13}$CO (2$-$1)    &$^{12}$CO (3$-$2)  &$^{13}$CO(3$-$2)   \\ \hline \hline
        RA (ICRS)\tablenotemark{$^{a}$}                    & 12:39:46.137       & 12:39:46.135      & $-$0.003                  & $-$0.004             & $+$0.001          &  $-$0.001          \\
        DEC (ICRS)\tablenotemark{$^{a}$}                   & $-$49.11.55.844    & $-$49.11.55.821   & $+$0.019                  &$-$0.026              &$-$0.002           &  $-$0.016         \\
        Major axis ($\arcsec$)\tablenotemark{$^{b}$}&  0.77$\pm$0.04     &  0.79$\pm$0.04     & $<$0.47                 &  0.35$\pm$0.09         &  0.49$\pm$0.04    &  $<$0.46\\
        Minor axis ($\arcsec$)\tablenotemark{$^{b}$}&  0.21$\pm$0.07      &  0.20$\pm$0.03     & $<$0.097                 &  0.22$\pm$0.11        &  0.21$\pm$0.04    &  $<$0.14\\
        Position Angle ($^{\circ}$)\tablenotemark{$^{c}$} &154$\pm$3     &  159$\pm$2         &  -                        & 119$\pm$59         &  158~$\pm$~6      &  - \\
        Peak Intensity \tablenotemark{$^{d}$}       &  0.24$\pm$0.03    &  0.42$\pm$0.05    &  72~$\pm$~8           &  55~$\pm$~   7   & 102~$\pm$~11       &  87 ~$\pm$~11            \\
        Integrated  Flux \tablenotemark{$^{e}$}     &0.51$\pm$0.05       & 1.39$\pm$0.16     & 91~$\pm$~12           & 72~$\pm$~11       & 220~$\pm$~27      & 138~$\pm$~20  \\
        Inclination \tablenotemark{$^{f}$}          & 75$\pm$5       & 75$\pm$2        &   -                        &      -                &    -                &   -            \\
        \hline \hline
        Beam Properties and image RMS \\\hline
        Major axis (\arcsec)                             &       0.52         &       0.34        &  0.56                     &  0.59                 &  0.37             &  0.39    \\ 
        Minor axis (\arcsec)                             &       0.45         &       0.29        &  0.48                     &  0.51                 &  0.31             &  0.33    \\ 
        Position Angle($^{\circ}$) (\arcsec)              &       68.1         &       73.9        &  69.9                     &  70.5                 &  73.3             &  74.3   \\  
        RMS (mJy~beam$^{-1}$)                                   &       0.008        &       0.017       &  0.79                     &  0.73                 &  1.11             &  1.53    \\  
        RMS Moment 0 (mJy~beam$^{-1}$~km~s$^{-1}$)               &      -            &           -        &  3.6                      &  3.9                 &  4.8              &  5.6   \\
        \hline
    \end{tabular}
   \tablenotetext{a}{Centroid coordinates for $^{12}$CO and $^{13}$CO are relative to the continuum centroid for the respective band, in units of arcseconds.}
   \tablenotetext{b}{FWHM deconvolved from the beam. '<' indicates the Gaussian component is a point source, and the major and minor axis sizes listed are upper limits.}
   \tablenotetext{c}{Position Angle of the deconvolved Gaussian component derived from IMFIT (measured from north through east)}.
   \tablenotetext{d}{Units are mJy~beam$^{-1}$ for continuum and mJy beam$^{-1}$ km s$^{-1}$ for moment 0 images. }
   \tablenotetext{e}{Units are mJy for continuum and mJy~km s$^{-1}$ for line images. }
   \tablenotetext{f}{Calculated with the  arccos of  minor axis divided by major axis, both measured by \sc{IMFIT}.}
  \end{center}
\end{table}

\subsection{Continuum and line images}\label{contobs}

Figure~\ref{cont_images} shows the 1.3 and 0.8~mm continuum images of
the edge-on disk around HD~110058. The disk is resolved in both
images.  The radial surface brightness profiles (averaged in the
direction perpendicular to the major axis) of the continuum images are
shown in the right panels of Figure~\ref{band6-co} and
Figure~\ref{band7-co}.  The band~7 image shows a 1-dimensional slice
along the disk's major axis. The surface brightness profile 
suggests the two peaks of an edge-on ring are marginally resolved.
The northwest peak appears brighter than the one located
towards the southeast, but not at a significant level. The band 6
continuum  profile peaks at the star's position, without
any notable features, due to the lower spatial resolution.

The total integrated continuum fluxes at 1.32 and 0.88~mm are 0.51$\pm$0.05~mJy and
and 1.39$\pm$0.16~mJy respectively, integrated within a $1.0$\arcsec radius
centered at the position of the star. The error bars in the derived
fluxes include the statistical errors and ALMA's 10\%
nominal absolute flux calibration accuracy in bands 6 and 7 (see
Table~\ref{table-1}). The spectral index between the two frequencies
is 2.5~\ah{$\pm0.4$}, consistent with the flux reported at 239.0~GHz by
\cite{Lieman2016} within uncertainties.

We use the {\sc CASA} task {\tt imfit} to fit a 2D Gaussian to the
continuum data.  Table~\ref{table-1} presents the source properties
derived from the Gaussian fitting at each frequency. The disk is
resolved in both directions in both bands. The deconvolved Gaussian
size in the direction of the major axis of the disk is 0.77\arcsec
($\sim$100~au), while the deconvolved size perpendicular to this is
0.2\arcsec~($\sim$26~au). The inclination of the disk derived from the
ratio of the minor and major axis (assuming a flat circular disk/ring)
is $\sim$75$^{\circ}$ in both bands. This inclination is quite
different from the near-IR images showing a disk very close to an
edge-on orientation. In Section~\ref{modeldust} we do a more formal
derivation of the disk's inclination, and discuss its implications in
Section~\ref{sec:vertical_dust}.

Figure~\ref{band6-co} and Figure~\ref{band7-co} show the moment 0
(contours) and moment 1 (colorscale) images for the observed
transitions of $^{12}$CO and $^{13}$CO. The disk is detected in all
lines. The source properties are listed in Table~\ref{table-1}.
Integrated fluxes are reported in Table~\ref{table-1}, and were
integrated within a circular region centered at the position of the
star of $0.65\arcsec$ in radius. The $^{12}$CO(2$-$1) integrated flux
of 90.6~$\pm$~12.0~mJy~km s$^{-1}$ is consistent with the values
reported in previous, lower-resolution, ALMA observations
\cite[92~$\pm$~17~mJy~km~s$^{-1}$, ][]{Lieman2016}. The moment 1 maps
show the typical velocity field of a disk in Keplerian rotation,
which is also clear in the position-velocity (PV) diagrams (see
Appendix~\ref{pvdiag}). Similarly to \cite{matra2017a}, we also
computed PV for the line ratios, and also for the consequential
optical depth using the (2-1) and (3-2) transitions
separately. These are also shown in Appendix~\ref{pvdiag}.  The
integrated spectra for both transitions of both isotopologues, and
the derived optical depths per channel, are also shown in the
appendix. The resulting optical depths \ah{\citep[calculated assuming an
interstellar 12C/13C abundance ratio of 76; ][]{Wilson1994}} indicate that the disk
is optically thick in both transitions. This is consistent with the
fact that the $^{12}$CO and $^{13}$CO fluxes are roughly the same,
suggesting the $^{12}$CO emission is highly optically thick. Also,
the ratio of the (3-2) to (2-1) fluxes goes roughly as $\nu^2$,
which also suggests the emission is optically thick.

The 1-dimensional radial surface brightness profiles of $^{12}$CO and
$^{13}$CO are shown in Figure~\ref{band6-co} and
Figure~\ref{band7-co}. Compared to the dust disk, the gas disk is more
compact.

The 2D Gaussian fit of the moment 0 maps from {\tt imfit} suggests
that the gas disk is marginally resolved along the minor axis in
$^{13}$CO(2-1) and $^{12}$CO(3-2). This would indicate that this disk
is not perfectly edge-on, and its inclination is slightly lower than
$90^{\circ}$.
To verify whether the
$^{12}$CO(3-2) disk is truly resolved in the direction of the disk's
minor axis we computed normalized intensity profiles in the direction
of the disk's minor axis. We compared the FWHM of the profiles to the
FWHM of the projected beam size, and conclude that the $^{12}$CO(3-2)
gas disk is only marginally resolved in the direction of the minor
axis. We, therefore, \ah{refrain from using} the moment 0 maps to obtain
information on the basic gas disk properties, and instead in
Section~\ref{modelgas} we use full radiative transfer modeling to
obtain the disk's parameters such as inclination and radius.

\section{Radiative transfer models}\label{rtmodel}

In this section, we use radiative transfer codes to fit the continuum and spectral line visibilities in order to constrain the distribution of dust and gas in this system. To fit the continuum data, we use the python package {\sc disc2radmc}\footnote{https://github.com/SebaMarino/disc2radmc} \citep{marino2022} that allows to create disk models and uses  {\sc RADMC-3D} \footnote{http://www.ita.uni-heidelberg.de/~dullemond/software/radmc-3d} \citep{Dullemond2012} to compute synthetic images. These images are then used to calculate model visibilities and a $\chi^2$ as in \citet{marino2018}. Note that we rescale the visibility weights of each band separately by a factor such that the reduced $\chi^2$ of our best fit model is equal to 1 \citep{marino2021}. This is to ensure the uncertainty estimates are correct. \footnote{\ah{While the relative weights/uncertainty of the visibilities are well
  estimated after the standard calibration of the visibilities, their
  absolute value tends to be off by a small factor between 1 and 2
  \citep[e.g. ][]{marino2018,matra2019}. This offset does not affect the imaging process (hence why it is not
  generally considered), but it does affect the derived parameter
  uncertainties as it changes the $\chi^2$.}} The line data is modeled using the {\sc pdspy} code from \citet{Sheehan2019}.

\subsection{Dust ring  model}\label{modeldust}

\begin{deluxetable*}{lcl}
\tablecaption{Best fit parameters of the ALMA continuum data using the parametric model in \S\ref{modeldust}.}
\tablehead{
\colhead{Parameter} & 
\colhead{Best fit value} & 
\colhead{description} \\
}
\startdata
$M_\mathrm{dust}$ [\mearth] & $0.080_{-0.003}^{+0.002}$  & Total dust mass \\
$r_{c}$ [au] & $31_{-8}^{+10}$  & Disk peak radius  \\
$\Delta r_{\rm in}$  [au]& $26_{-18}^{+26}$  & Inner FWHM  \\
$\Delta r_{\rm out}$  [au]& $72_{-12}^{+9}$  & Outer FWHM  \\
$h$  & $0.17^{+0.05}_{-0.09}$  & Vertical aspect ratio  \\
$i$ [$^{\circ}$] & $78^{+7}_{-3}$  & Disk inclination from face-on \\
PA [$^{\circ}$ ]& $157\pm1$ & Disk position angle \\
$\alpha_\mathrm{mm}$ & $2.45\pm  0.06$ & Spectral index 
\enddata
\tablecomments{The values correspond to the median, with uncertainties based on the 16th and 84th percentiles of the marginalized distributions.}
\label{rtdusttable}
\end{deluxetable*}

The dust disk is modeled as an axisymmetric ring with a surface density following an asymmetric Gaussian distribution
\begin{align}
\label{eq:mycase2}
\Sigma_{\rm dust}(r) = \Sigma_c  \begin{cases}
\exp[-\frac{(r-r_c)^2}{2\sigma_{\rm in}^2}] & \text{if $r\le r_c$}, \\
\exp[-\frac{(r-r_c)^2}{2\sigma_{\rm out}^2}] & \text{if $r> r_c$},
\end{cases}
\end{align}
where $\sigma_{\rm in, out}$ are the standard deviations interior and exterior to the surface density peak at $r_c$. Vertically, the disk is assumed to have a Gaussian distribution with a standard deviation $H(r)$, which is equivalent to assuming a Rayleigh distribution of inclinations \citep{matra2019}. The model free parameters are the total dust
mass ($M_{\rm d}$), the peak radius ($r_c$), the inner and outer full width half maxima
($\Delta r_{\rm in, out}$)\footnote{The true full width half maximum is $(\Delta r_{\rm in}+\Delta r_{\rm out})/2$}, its vertical aspect ratio ($h$, where $h=H/r$ is constant across the disk), the disk
inclination ($i$), its position angle (PA), the disk spectral index ($\alpha_{\rm mm}$) and phase centre offsets for both band 6 and 7 observations. We leave $h$ as a free
parameter since this disk is highly inclined and the minor axis resolved, and thus these
observations could provide good constraints. We use uniform priors for
all the free parameters, restricting $h$ to values in the range
$0.015-0.4$ for computational reasons. The dust grain properties are
the same as in \citet{marino2018}, dust species with a mass weighted
opacity assuming a size distribution from 1~$\mu$m to 1~cm, with a power-law
index of -3.5, and made of a mix of astrosilicates, amorphous carbon
and water ice. This choice only has an effect on the dust opacity and
derived mass, but not on the dust distribution. The stellar radius was
fixed to 1.6~ R$_{\odot}$ and the stellar temperature to 8000~K,
consistent with the system's parameters \citep{saffe2021}. These parameters determine the
dust temperature (calculated with {\sc RADMC-3D}) and stellar flux in bands 6
and 7 (2~$\mu$Jy and 5~$\mu$Jy, respectively, below our observations'
detection limit).

\begin{figure}[h!]
    \centering
    \includegraphics[width=0.6\textwidth]{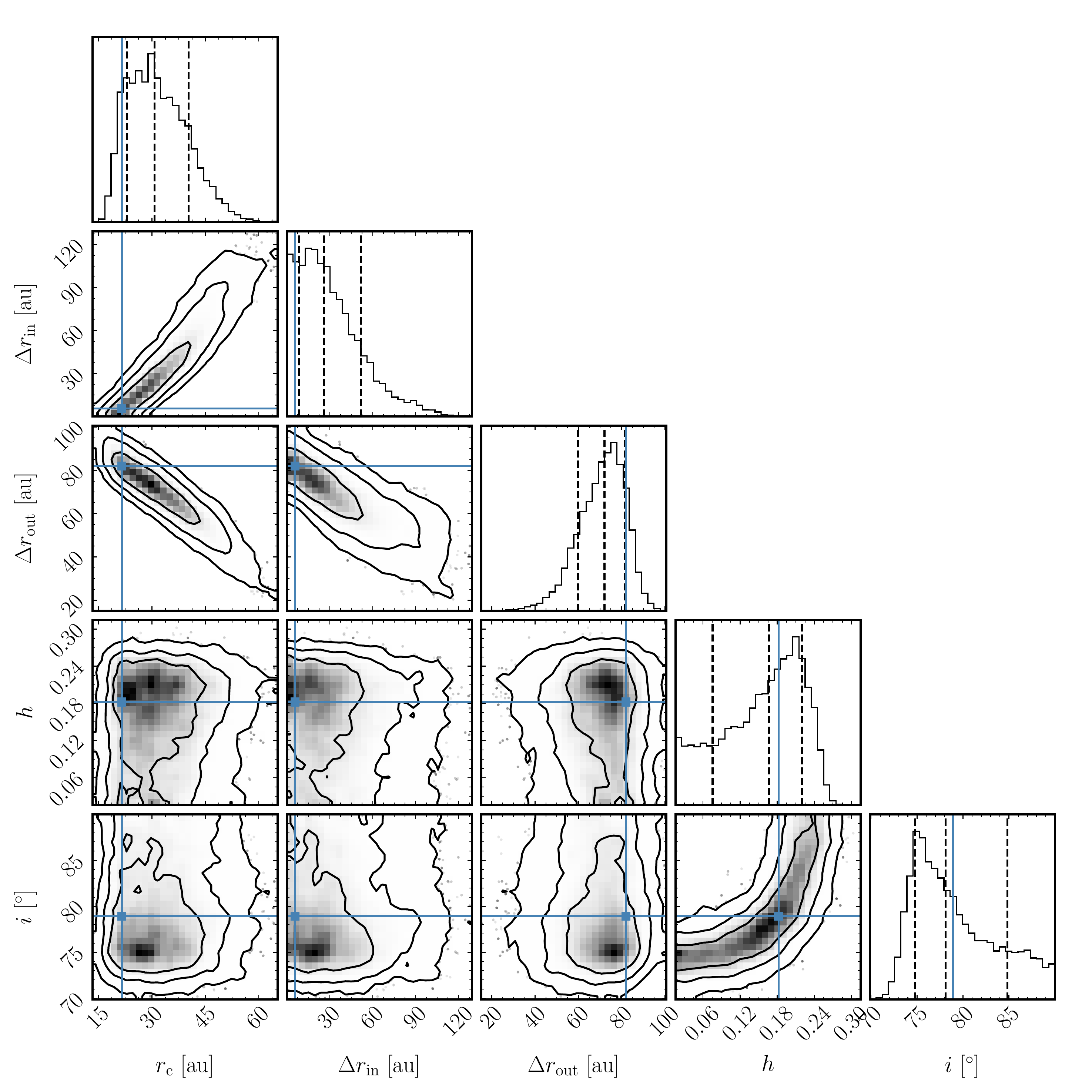}
    \caption{Posterior distribution obtained for the disk peak radius, inner and outer FWHM, vertical aspect ratio, and inclination. The blue lines show the parameter values that give the best fit, whereas the vertical dashed lines show the 16th, 50th and 84th percentiles. The contours represent the 68, 95 and 99.7\% confidence levels.  }
        \label{fig:corner}
\end{figure}

\begin{figure}[h!]
    \centering
    \includegraphics[scale=0.2]{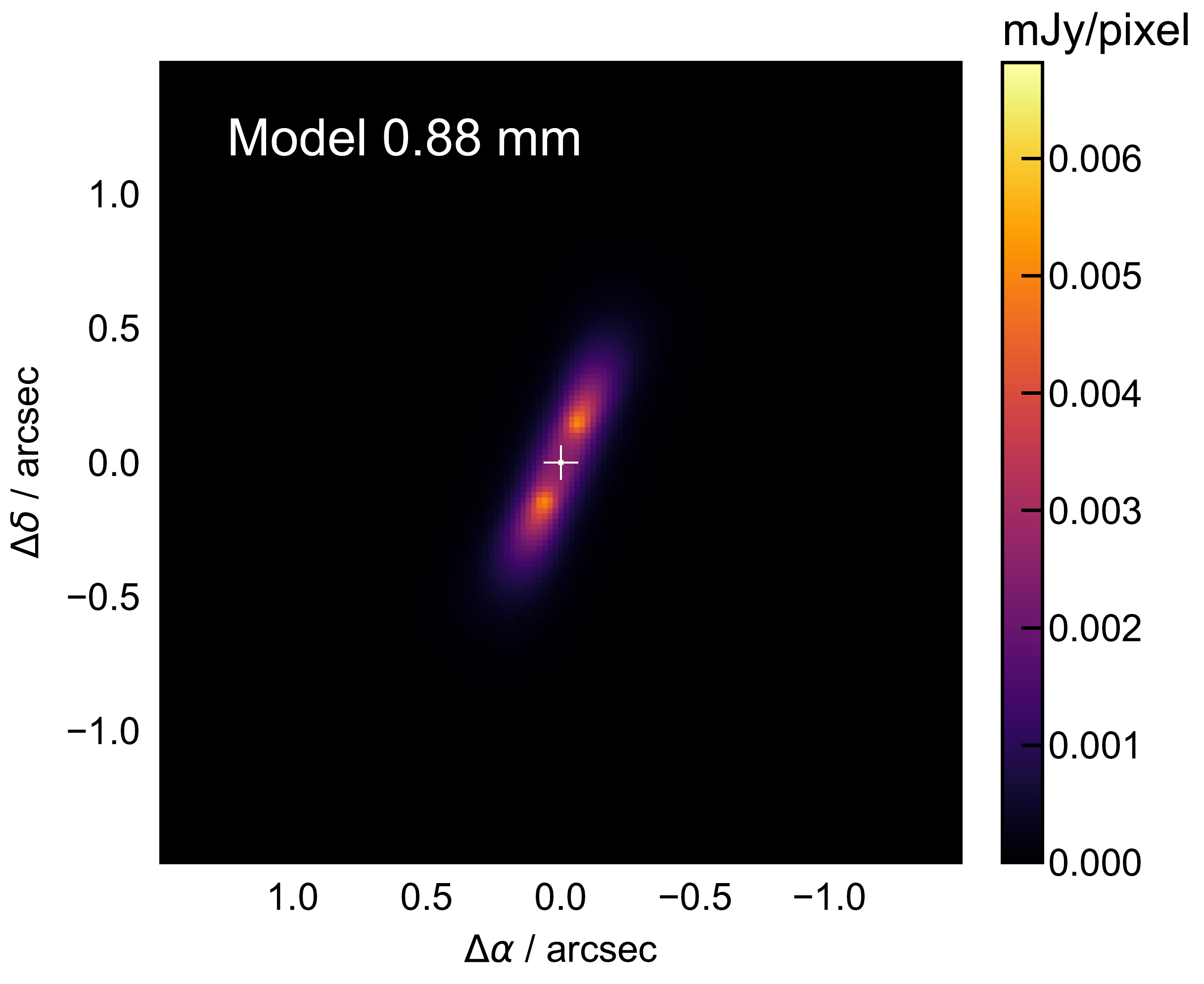}
    \includegraphics[scale=0.2]{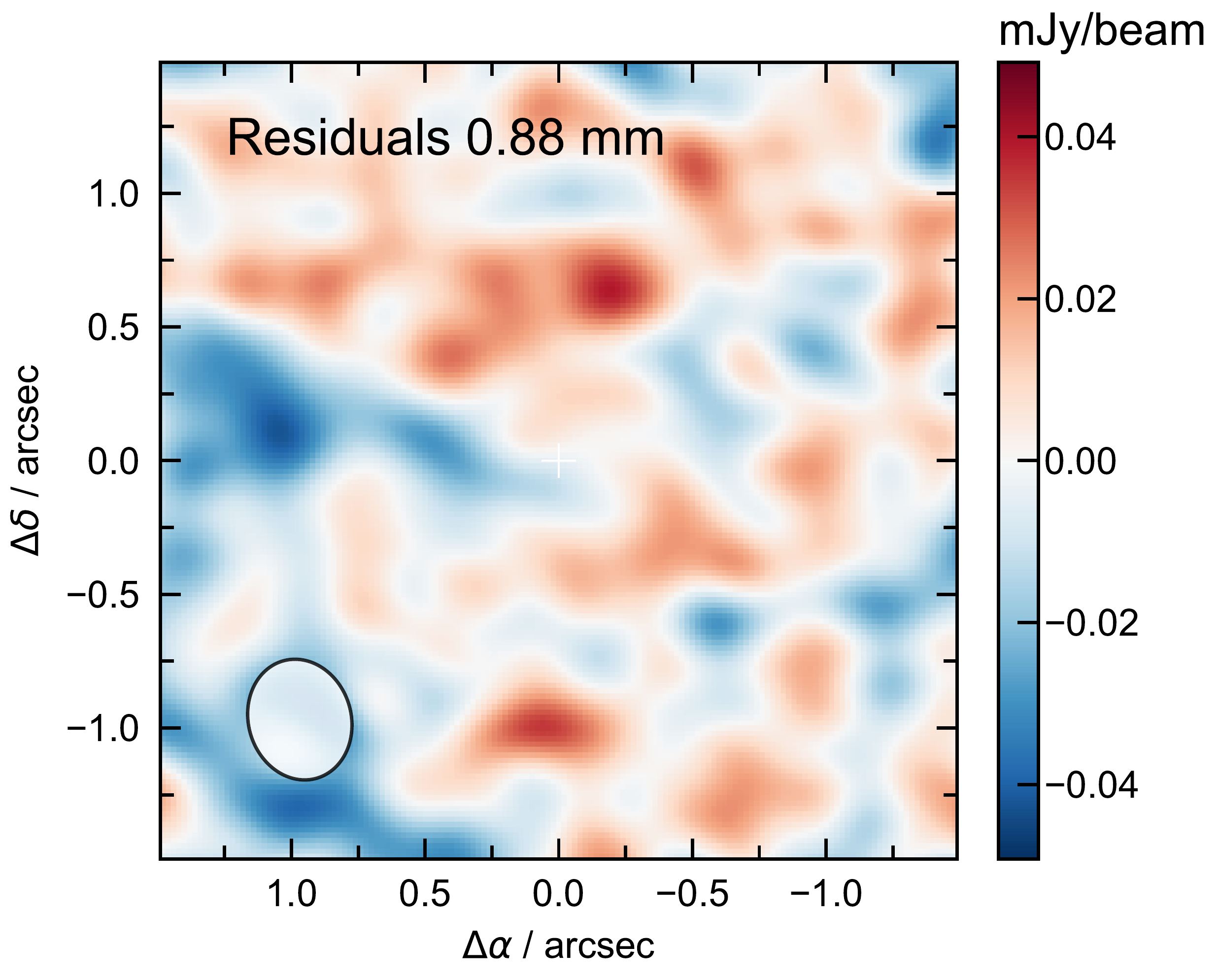}
    \includegraphics[scale=0.2]{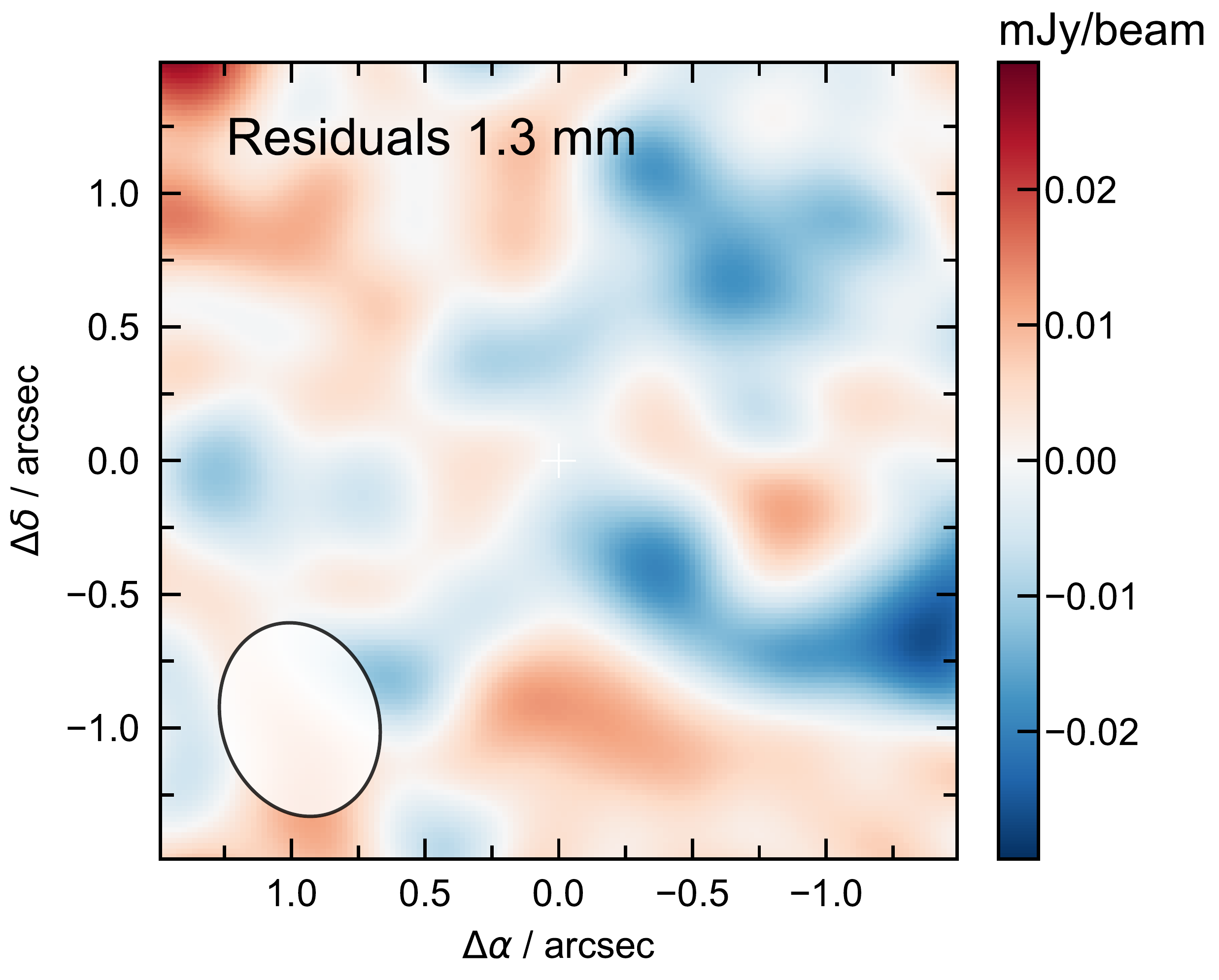}
    \caption{ {\bf{Left:}} Band 7 best-fit dust ring  model image. {\bf{Center:}} Band 7 residual image after subtracting the model to the data. Color stretch ranges from -3$\sigma$ to 3$\sigma$, where $\sigma$ = RMS of 1.7$\times 10^{-2}$ mJy~beam$^{-1}$.  {\bf{Right:}} Band 6 residual image after subtracting the model to the data. Color stretch ranges from -3$\sigma$ to 3$\sigma$, where $\sigma$ = RMS of 8.6$\times 10^{-3}$ mJy~beam$^{-1}$. }
        \label{band6-model}
\end{figure}

The parameter space is explored using a Markov Chain Monte Carlo
(MCMC) routine. For each set of parameter values, {\sc disc2radmc} is
used to compute the dust density distribution and {\sc RADMC-3D} to
compute a synthetic image at 0.88~mm, which is then used to calculate an
image at 1.3~mm based on a uniform disk spectral index $\alpha_{\rm
  mm}$ that is left as a free parameter 
 These model images are
multiplied by the corresponding ALMA primary beam, and finally fourier
transformed to produce model visibilities that can then be compared to
the data. The Band 6 and Band 7 data were fitted simultaneously using
the full aggregate bandwidth from all line-free channels. The posterior distribution
is constrained using the Goodman \& Weare's Affine
invariant MCMC Ensemble Sampler in the {\sc emcee} code
\citep{foreman2013}. The best-fit parameters for the dust ring model
are presented in Table~\ref{rtdusttable} and the posterior
distribution of the most relevant parameters in
Figure~\ref{fig:corner}.

These results indicate the disk is centered at $31^{+10}_{-8}$~au with an inner edge that is sharper than the outer edge (i.e. $\Delta r_{\rm in}<\Delta r_{\rm out}$). Defining the disk inner edge location as r$_{c}$-$\Delta
r_{\rm in}$/2, and outer edge location as r$_{c}$+$\Delta r_{\rm out}$/2, we find that the disk
inner edge is at $18_{-4}^{+3}$~au  (smaller than 23 au at 99.7\% confidence) and its outer edge is at $67\pm4$~au (larger than 59 au
at a 99.7\% confidence).
Since the mm-sized dust traces the
planetesimal distribution, these observations thus reveal that the
planetesimals are distributed in a 49~au wide belt with a fractional width (disk width over its disk centre) of 1.2 or 1.6 depending on the definition of its disk centre. This is anyway much higher than the median fractional width of debris disks observed with ALMA of 0.7 (Matrà et al. in prep). The derived outer radius of ${\sim}70$~au is consistent with the
estimate from \citet{Lieman2016}. The inner radius of the dust ring
is inferred for the first time, although its exact value is likely
dependent on the assumption of a Gaussian density distribution. Higher
resolution observations would be needed to determine this with
confidence. \ah{Nevertheless, as part of a separate project (Terrill et al. in prep), we explored that when using a non-parametric model we find a radial profile that peaks at around 30~au, with a smooth outer edge. This is consistent with our model choice of a simple Gaussian profile, allowing for the inner and outer regions to have different widths/standard deviations.}

Consistent with scattered light observations, the disk is found to be
very inclined although the marginalized distribution is still
consistent with a wide range of values from $90^{\circ}$ (perfectly
edge-on) down to 71$^{\circ}$ (99.7\% lower limit). This behavior is
due to the disk being resolved along the minor-axis, thus if the disk
is vertically thin (low $h$), the inclination must be below
${\sim}80^{\circ}$ to account for the minor-axis width. This also explains
the correlation between $h$ and the inclination in
Figure~\ref{fig:corner}, with high values of $h$. Interestingly,
scattered light observations and our gas modelling presented below
suggest the disk to be very inclined ($i>80^{\circ}$), which would
suggest the disk is vertically resolved with $h=0.13-0.28$ (99.7\%
confidence interval). In \S\ref{sec:vertical_dust} we discuss this
finding and its implications.

\subsection{Gas disk model}\label{modelgas}

To provide more formal constraints on the gas disk properties, we use
the {\sc pdspy} code from \citet{Sheehan2019} to fit the $^{12}$CO and
$^{13}$CO emission. The code generates synthetic line emission maps
that can be readily compared to the data. It uses {\sc RADMC-3D}   to
produce synthetic line emission maps which are then sampled similarly
to the visibility data using the fast sampling code {\sc GALARIO} \citep{tazzari2017}.
The synthetic model visibilities are compared to the data using a Bayesian
approach in which the probability distribution is sampled via Markov
Chain Monte Carlo (MCMC) method implemented in the {\sc emcee} code.

The model assumes a passively irradiated disk in hydrostatic
equilibrium rotating with a Keplerian velocity field (the vertical and radial velocities are zero).  The radial
surface density of the disk is given by the standard \citet{LBP1974}
profile, which corresponds to a power-law disk with an exponentially
decaying tail at large radii,

\begin{eqnarray}
  \Sigma(r)=\Sigma_{\rm 0} \left( \frac{r}{R_{\rm c}} \right)^{-\gamma} \exp \left[- \left( \frac{r}{R_{\rm c}}\right)^{2-\gamma} \right].
\end{eqnarray}\label{tapered}

\noindent The characteristic radius R$_{\rm c}$ represents the radius
where the exponential tail starts to dominate the density profile, and is a proxy for the outer radius of the disk.
 The power-law surface density exponent $\gamma$ also  controls
how sharply the disk is truncated in the exponential
tail. The reference surface density $\Sigma_{\rm 0}$ is related to the total disk mass M$_{\rm
  disk}$ as

\begin{equation} 
M_{\rm disk} = \frac{2 \pi R_c^2 \Sigma_{\rm 0} }{2 - \gamma}.
\end{equation}

\noindent We assume that the disk is vertically isothermal, with the radial temperature distribution of the gas is defined by a power-law  \begin{equation}\label{eqn7}
  T(r)=T_0\Big(\frac{r}{1\ \mathrm{au}}\Big)^{-q}. \end{equation} Solving for
hydrostatic equilibrium, the vertical scale height as a function of
radius is 

\begin{equation}
H(r) = \left[ \frac{k_B\, r^3\, T(r)}{G\, M_*\, \mu m_H} \right]^{1/2},
\end{equation}

where M$_*$ is the mass of the central star, $\mu$ is the mean
molecular weight of the gas, and k$_B$ and G are Boltzmann's and
gravitational constants, respectively. The code considers local thermodynamic equilibrium (LTE) to generate the images, which is valid if the gas densities are high enough. We assume $\mu=14$, which
corresponds to the case where the gas is predominantly composed of
carbon and oxygen atoms released by photodissociation of CO
\citep{kral2016}.

The model adopted here includes the following free parameters:
total disk mass M$_{\rm disk}$, stellar mass M$_{*}$, disk
characteristic radius $R_{\rm C}$, disk inner radius $R_{\rm in}$ (inside
of that location the density drops to zero), T$_0$ (the temperature at
1~au), position angle PA, the surface density power law exponent
$\gamma$, the system's radial velocity v$_{sys}$ (LSRK), and offset from the phase
center x$_0$ and y$_0$. The dynamical mass of the central object
M$_{*}$ is fitted assuming the distance of 129.9~pc \citep{gaia2018}.
The radial exponent for the temperature dependence $q$ is fixed to
0.5. The $^{13}$CO isotopologue
ratio with respect to $^{12}$CO was set to the canonical value of 76
\citep[][]{Wilson1994}.

We chose to fit the $^{12}$CO(3-2), $^{13}$CO(3-2) and $^{13}$CO(2-1)
data simultaneously to exploit the higher angular resolution of the Band
7 data while still using the (2-1) lines of $^{13}$CO to solve for the
temperature/mass degeneracy. We found that adding the extra
$^{12}$CO(2-1) made the running time of the fit prohibitively large.

\begin{figure}
\begin{center}
\epsscale{0.5}
\includegraphics[angle=0,scale=0.3]{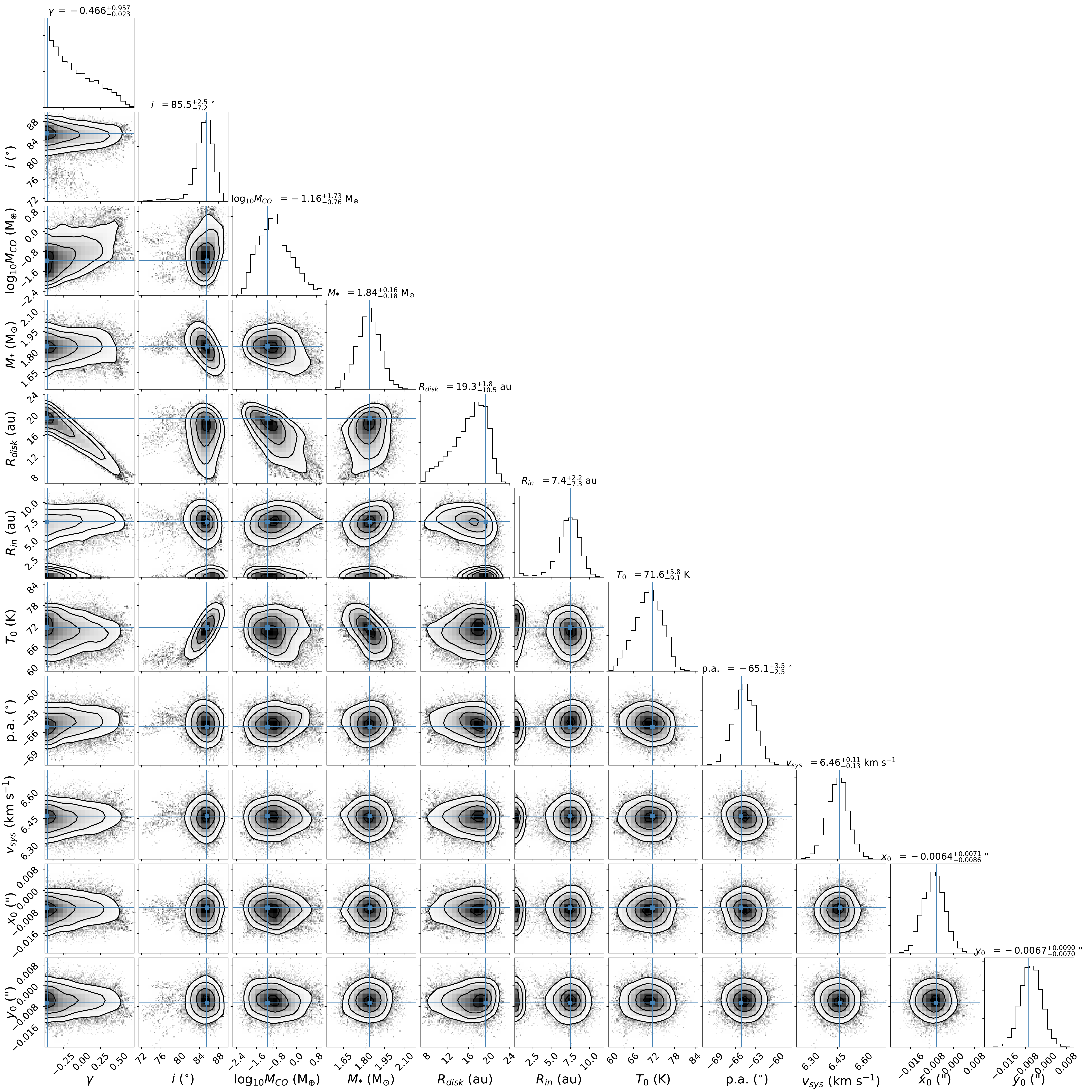}

\caption{Triangle plots of the posterior probability distribution
  function for the gas disk model that fits $^{12}$CO(3-2),
  $^{13}$CO(3-2), and $^{13}$CO(2-1) simultaneously. The blue lines
  show the point with the highest probability, i.e. the most likely
  model from the fit. The maximum likelihood values and
  uncertainties corresponding to the range around these values including 95\% of all walkers in the posterior are shown at the top of each column.}
\label{fig:cofit}
\end{center}
\end{figure}

The MCMC run started with 100 walkers, which were left to evolve
during 1100 steps for the burn-in phase. Following this burn-in phase,
the MCMC code ran for another 2000 iterations to sample
the posterior probability distribution. The resulting best-fit parameters and uncertainties are presented in Table~\ref{rcotable}. We report the maximum likelihood model from our fit as the best-fit parameters, and use the range around the best fit parameter values including 95\% of the posterior samples to report the uncertainties on these values. We choose to report the 95\% confidence intervals here because the posteriors for some parameters from the fit are highly skewed, with the maximum likelihood value falling outside of the 68\% confidence interval. The marginalised probability distributions are presented in Figure~\ref{fig:cofit}.

\begin{figure}
\begin{center}
\epsscale{0.5}
\includegraphics[angle=0,scale=0.8]{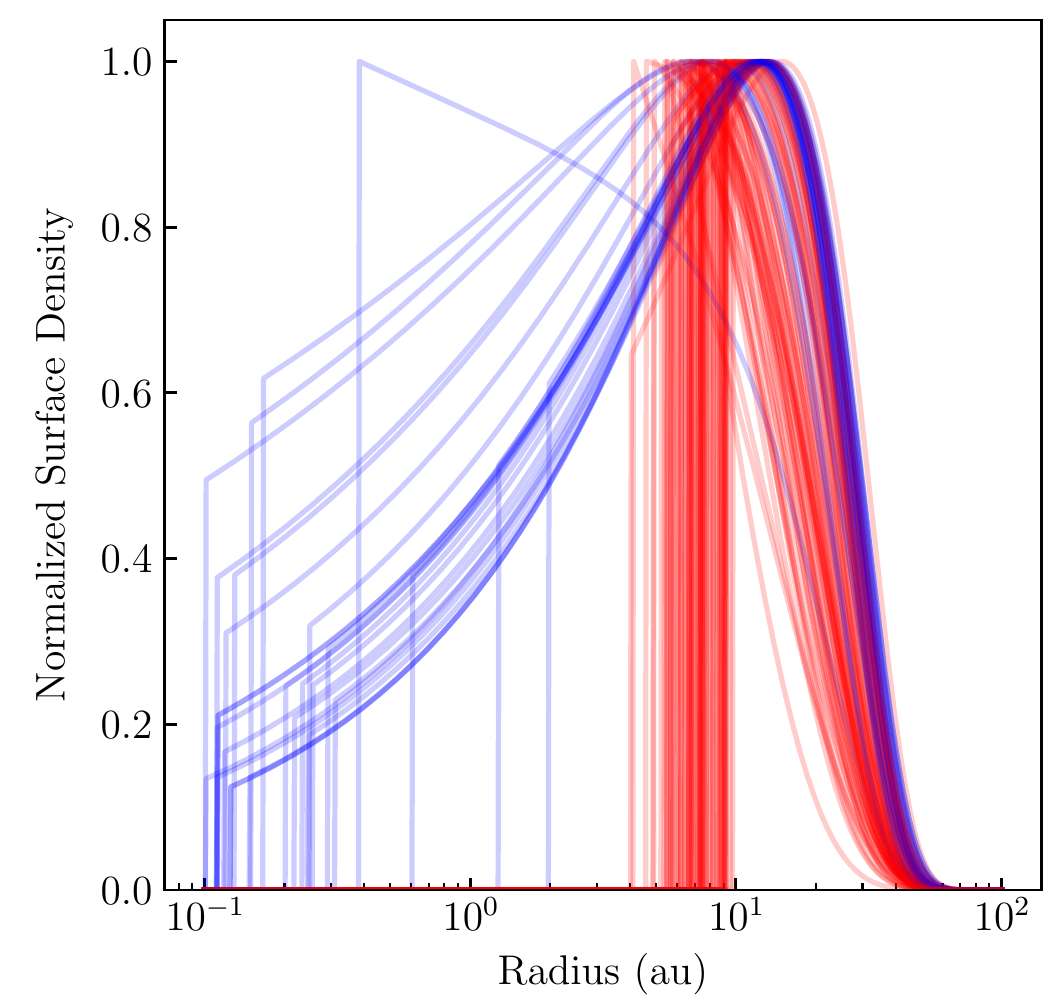}

\caption{Normalized CO surface density distribution from the MCMC
  run. Coloring is based on R$_{in}$ $ <$ 3 au (blue) and R$_{in} >
  $3 au (red) from the two 'best-fit' solutions.}
\label{fig:surfacedesity}
\end{center}
\end{figure}

The disk inclination of \ah{$ 85.5_{-7.2}^{+2.5} $} determined for the gas disk is compatible with
the disk being very close to edge-on. It is also consistent with the
inclination derived for the mm dust disk. The stellar mass
of 1.84$_{-0.18}^{+0.16}$ M$_{\odot}$ is well constrained, with
minimal dispersion, and is in better agreement with the star being late A- dwarf, closer to A6/7. This revised stellar classification seems consistent with recent derivation of the stellar temperature of T$\sim7839\pm202$~K from optical spectroscopy \citep{saffe2021}, which also suggest that HD~110058 is an A6/7V star rather than A0V\footnote{\tt{ https://www.pas.rochester.edu/$\sim$emamajek/EEM\_dwarf\_UBVIJHK\_colors\_Teff.txt}} \citep{pecaut2013}. The
determined system's radial velocity of $6.4~\pm0.1$~km~s$^{-1}$ is consistent
with the heliocentric velocity of $\sim$ 12.6~km~s$^{-1}$ measured in
the optical \citep{hales2017,Rebollido2018} after converting to LSRK \ah{(5.6~$\pm2.4$ ~km~s$^{-1}$)} \footnote{\ah{The conversion from Heliocentric to LSRK velocity frame was computed using the RV software \citep{wallace1997} from the Starlink Software Collection \citep{currie2014}. }}.  The position angle reported by the fit uses the {\sc RADMC-3D} convention, i.e., the position angle of the angular momentum vector of the disk on the plane of the sky, which is offset by 90~degrees from
the position angle defined by the disk's major axis.
\citep[see][]{Czekala2019}. Therefore the result for the gas position
angle in the traditional convention is 155.1$_{-2.5}^{+3.5}$$^{\circ}$, in
agreement with the $157\pm1$$^{\circ}$ position angle derived for the mm dust disk (Section~\ref{modeldust}), and the $155\pm1$$^{\circ}$ measured in scattered light \citep{kasper2015}.

The most relevant parameters to constrain the origin of the carbon
monoxide present in the disk are the disk's inner and outer radius,
the total CO mass and the resulting surface density distribution. The
total CO mass present in the disk as derived from the fitting of the
$^{12}$CO(3-2), $^{13}$CO(3-2) and $^{12}$CO(2-1) data is 
$0.069^{+3.56}_{-0.057}$ \mearth.
{Even taking the lower end of the 95\% confidence interval, this is more than two orders of magnitude larger than the CO mass estimated by \citep{kral2017} using a disk model that accounts for optical thickness and non-LTE, and three orders of magnitude larger than CO mass derivations based on previous $^{12}$CO(2-1) data that assume optically thin emission \citep[e.g., ][]{moor2017}. We note, however, that this mass is derived under the assumptions of the model that we employ, and this difference could be in part due to systematic effects related to our relatively simplistic choice of model. For example, we assume the disk is vertically isothermal, though this is almost certainly not the case, in LTE, and assumes that the surface density profile can be well approximated by a tapered power law. If any of these assumptions are incorrect, it could have an effect on the mass we derive. Including such additional physics \citep[e.g., ][]{Rosenfeld2013} and alternative surface density distributions in future modeling could potentially alter this picture.

The disk's characteristic radius R$_{\rm c}$ derived from the fit is
19.3$_{-10.5}^{+1.8}$~au, which confirms that the gas disk is more
compact than the dust disk that has an outer edge of 70~au. We discuss this difference in \S\ref{sec:distribution_gas}.

We find that $R_{\rm in}$ is not well constrained, showing a bimodal distribution corresponding to two solutions: $R_{\rm in}$ smaller or larger than  ${\sim}3$~au. \ah{That said, while the inner edge of the disk itself is not well constrained, we do find that these two families of solutions provide a consistent picture of where the CO gas is located.} Figure~\ref{fig:surfacedesity} presents the normalized surface density profiles for a random sample of the posterior distribution, colour-coded by $R_{\rm in}$: blue for $R_{\rm in}<3$~au and red otherwise. From this figure, we can see how most solutions with small $R_{\rm in}$ (blue) have surface densities that increase with radius up to $R_{\rm c}$, effectively creating a cavity depleted of CO gas. Solutions with a large $R_{\rm in}$ tend to converge to $R_{\rm in}\sim7$~au. 
We also note that though the posterior for R$_{in}$ is bimodal, this could in part be due to our choice of prior on $\gamma$. $\gamma$ controls the surface density profile, with negative values of $\gamma$ leading to the smoothly increasing surface density profiles for the solution with $R_{\rm in} < 3$~au. More negative values of $\gamma$ lead to more sharply increasing surface density profiles that more closely approximate the truncated disk solution, which could in turn plausibly have larger values of $R_{\rm in}$. In our modeling, we place a limit of $\gamma > -0.5$, however lowering this limit could potentially thereby fill in the gap between the two modes of the posterior that we see.

\ah{Hence, although we cannot place a strong constraint on where the formal inner edge of the disk is, both solutions suggest that the CO surface density profile peaks near $\sim10$ au and has a depletion of material within that radius. We further note that} gas at radial distances smaller than 10~au should result in gas emission at velocities larger than 10~km~s$^{-1}$, but such emission is not visible at a significant level in Position-Velocity diagrams of the data nor of the best-fit model (see Appendix A).  
Higher spatial resolution and sensitivity data is required to determine in more detail the surface density profile of the CO gas.

We note that the temperature at 1 au (T$_0$) found by the modeling, of $71^{+5.7}_{-9.1}$ K, is seemingly quite small for the proximity to an A0 star. That said, 1 au is well below the limits of our observations, and as Figure \ref{fig:surfacedesity} shows, we also find a dearth of material at that radius. As such, we suspect that the low temperature is likely the result of the model matching the temperature at larger radii that are probed by our observations, and extrapolating inwards using the fixed temperature power-law exponent ($q = 0.5$), rather than a true estimate of the disk temperature at 1 au. 
We note that the temperature at 1 au ($T_0$) corresponds to a temperature of $\sim16$ K at 20 au where the gas density peaks. This temperature seems low for the proximity to an A-type star. This may, however, once again be due to our relatively simple choice of model (or if the gas is in non-LTE and $T$ different from $T_{\rm ex}$). The temperature controls the height of the disk, and so the low temperature may be required to match the geometric profile of the disk.  A more physically motivated model with a cold disk midplane and a warm atmosphere \citep[e.g., ][]{Rosenfeld2013}, or a CO vertical distribution skewed towards the midplane due to photodissociation \citep{marino2022}, may better match the expected temperature of the disk at this radius while also producing the proper height of the gas disk.

%
%
%
%
%
%
%

\begin{deluxetable*}{lcc}
\tablecaption{Best-fit Gas Disk Model Parameters  }
\tablehead{
\colhead{Parameter} & 
\colhead{Best fit value} & 
\colhead{description} \\
}
\startdata
$\gamma$ &                    $-0.47_{-0.02}^{+0.95} $       & disk CO surface density exponent \\ 
Incl [$^\circ$] &                        $ 85.5_{-7.2}^{+2.5} $        & disk inclination from face-on \\
$\log_{10}$(M$_{\rm CO}$) [\mearth]  &   $ -1.16_{-0.77}^{+1.72} $       & total CO mass \\
M$_{\rm star}$ [M$_{\odot}$ &              $ 1.84_{-0.18}^{+0.16} $        & Stellar Mass \\
R$_{\rm disk}$ [au] &              $ 19.3_{-10.5}^{+1.8} $         & Disk Characteristic Radius (R$_{\rm C}$)\\
R$_{\rm in}$ [au] &                $ 7.4_{-7.3}^{+2.2} $          & Disk inner Radius \\
T$_{\rm 0}$ [K]&                   $ 71.6_{-9.1}^{+5.7}$         & Temperature normalization, at 1 au \\
PA [$^\circ$] &                          $ -65.1_{-2.5}^{+3.5} $         & Position angle of the disk's angular momentum vector \\
v$_{\rm sys}$ [km~s$^{-1}$] &                 $ 6.46_{-0.13}^{+0.11} $        & Systemic Velocity in LSRK \\
 x$_{\rm 0}$ [\arcsec] &                $ -0.0064_{-0.0086}^{+0.0071} $ & X- Offset \\ 
 y$_{\rm 0}$ [\arcsec]&               $ -0.0067_{-0.0071}^{+0.0091}$  & Y- Offset \\
\enddata
\tablecomments{The best-fit values corresponds to points with the highest probability (i.e. the the maximum likelihood model from the fit) and the 95\% confidence interval around that point.}
\label{rcotable}
\end{deluxetable*}


\section{Discussion}\label{discussion}

In this section we discuss the results regarding the vertical extent of the disk derived in \S\ref{modeldust}, and the CO gas distribution derived in \S\ref{modelgas}.

\subsection{Dust radial distribution}
\label{sec:radial_dust}

The new ALMA data resolve the circumstellar dust around HD~110058 and
provide better constraints on the disk's parameters compared to
previous observations from \citet{Lieman2016} which had resolution of
$1.3\arcsec\times0.8\arcsec$ (a factor of 3-4 lower than our new observations). The
analysis of the new data strongly suggests that the dust (and thus
planetesimal) disk is very wide with a FWHM of 49~au and a peak radius of 31~au. We
note that this disk has a very small peak radius compared
to other bright debris disks around stars of similar luminosity observed with
ALMA \citep[expected central radius of
  $110\pm20$~au; ][]{matra2018}. The only disk that appears to be
similar is HD~121191, with a central radius of 52~au and width smaller
than 61~au \citep{kral2020gassurvey}. HD~110058's disk being smaller
could be an effect of its short age ($\sim17$~Myr old),
meaning that the inner regions would be less depleted than in older
systems due to collisional evolution. However, other young disks around similar luminosity stars
like the ones around $\beta$~Pic, HD 131488, HD~131835 are all
significantly larger \citep{matra2019, moor2017, kral2019}. Thus it
appears that this system is at the tail of the radius distribution for bright debris disks.
In scattered light the disk is also seen small with signal detected to
only about 65~au \citep{kasper2015, esposito2020}, location that is
consistent with our inferred disk outer edge.

The disk inner edge is 18~au ($<23$~au at 99.7\% confidence), which is the smallest inferred inner edge inferred for an exoKuiper belt with ALMA (Matr\`a et al. in prep). This has strong implications for the evolution of gas, since the equilibrium temperature would be higher than 100~K and thus other volatiles apart from CO would readily sublimate \citep[e.g. CO$_2$][]{collings2004}. CO$_2$ sublimation could trigger an increase in outgassing, and since this molecule quickly photodissociates into CO \citep{hudson1971}, this would enhance the CO gas production rate in the inner regions.  

\subsection{Dust vertical distribution}
\label{sec:vertical_dust}

Our observations strongly suggest that the disk is vertically thick,
with an aspect ratio in the range 0.13-0.28 (99.7\% confidence) if
$i>80^{\circ}$ (i.e. consistent with the gas and scattered light
observations). The vertical thickness of debris disks is a key
property that traces the inclination dispersion, and thus it contains
valuable information about the dynamical history of a system. So far,
this has been inferred for two edge-on disks \citep[][]{matra2019,
  daley2019} and a few less inclined disks \citep{marino2019,
  Kennedy2018}. These measurements of $h$ range between $0.02-0.09$,
which translate to inclination dispersions ($i_{\rm rms}$) of
$2-7^{\circ}$ \citep[$i_{\rm rms}=\sqrt{2}h$,][]{matra2019} that are
close to the inclination dispersion of the cold population of the
classical Kuiper belt
\citep[$\sim3^{\circ}$,][]{brown2001}. $\beta$~Pic is an exception to
this as it was found to be best fit with a double population that is
analogous to the classical Kuiper belt \citep{matra2019}. These
populations have inclination dispersions of $1^{\circ}$ and
$9^{\circ}$. For HD~110058 we concluded that the inclination
dispersion is likely in the range $11-23^{\circ}$, which would make it
the thicker disk known to date, but still lower or consistent with the
classical Kuiper belt's hot population and scattered disk
\citep[$20-30^{\circ}$][]{brown2001}.

\subsubsection{Scattering}
This high degree of orbital stirring revealed by the high $i_{\rm  rms}$ strongly suggests that this disk has been perturbed by
planets. The level of stirring is a combination of the mass and number of stirrers (since stirring is localized), which raise the relative velocities over time. Following a similar procedure to \cite{matra2019} we can estimate the minimum planet mass that could cause this stirring through close encounters. We first use their equation 10 to find that the relative velocities are in the range 3-7~km~s$^{-1}$ at the peak radius. In order to reach these relative velocities, the massive bodies stirring the disk should have escape velocities close or higher
to those values, and thus we find the minimum stirrer mass as

\begin{equation}
    M_{\rm s} = 0.03 M_{\oplus} \left(\frac{h}{0.13}\right)^{3} \left(\frac{r}{\rm 31\ au}\right)^{-3/2}\left(\frac{ M_{\star}}{1.8\ M_{\odot}}\right)^{3/2}\left(\frac{\rho}{\rm 4\ g\ cm^{-3}}\right)^{-1/2},
\end{equation}
where $\rho$ is its bulk density and $r$ the orbital radius. Therefore, the stirrer should at least be as massive as Mercury. 

However, using equations 12 and 14 from \citet{matra2019}, we find that a planet with this minimum mass would be unable to stir the orbits to the observed levels on its own within the age of the system (17~Myr versus $\sim$ 500~Gyr that it would take to stir the system to this level on its own), and thus an unrealistic number of these planets closely packed would be needed to stir the disk. More feasible is that the stirring was caused
by multiple widely spaced planets with a larger mass. The intersection of equations 12 and 14 in \citet{matra2019} sets the minimum planet mass for which stirring can be achieved by the age of the system and planets are spaced by $\sim14$~Hill radii, ensuring stability and also stirring in between their orbits. For HD~110058, we find that minimum mass is 15~$M_{\oplus}$. Such planets could be embedded in the disk and have cleared gaps \citep[$>5$~au,][]{Wisdom1980} that our observations are unable to resolve yet \citep{marino2018, marino2019, marino2020hd206,
  macgregor2019}. 
  
Alternatively, the high inclinations could be due to
a single massive planet near the disk inner edge. Such planet could
have scattered most of the original population while migrating, in which case the disk would be dominated today by a population similar to the Kuiper belt's scattered disk, but much more massive
\citep{duncan1997}. Our modelling indeed suggests that surface density beyond the peak radius decays smoothly with radius  as expected if the disk is highly stirred \citep{marino2021}. Higher resolution
observations could constrain better the location of the disk inner
edge and the surface density profile, and thus constrain better this
scenario.

\subsubsection{Secular perturbations}

A different mechanism that could explain the high $i_{\rm rms}$ is through secular interactions. If the disk was initially misaligned to an inner massive planet, the disk would have been forced to the same orbital plane as the planet \citep{Wyatt1999}. This has been suggested to explain the warp in $\beta$~Pic \citep[][]{mouillet1997, augereau2001}. As the disk particles interact with the planet, their inclinations precess and after one secular timescale they form a thick disk with a vertical height equal to twice the initial mutual inclination between the disk and planet. Thus the derived inclination dispersion of $11-23^{\circ}$ would suggest an initial misalignment of $6-12^{\circ}$. Interestingly, this system shows a tentative warp in its outer regions \citep{kasper2015}. Warps are expected during this type of evolution. As the secular timescale increases with radius for an internal perturber, beyond a certain radius ($a_w$) disk particles will have not precessed enough to be aligned with the orbit of the planets. The location where this happens thus can constrain the mass and semi-major axis of the perturbing planet. Using Equation~4 from \cite{dawson2011}, we find that a warm exo-Jupiter at 3-10~au that was born misaligned or evolved to a misaligned orbit could explain the warp location. If the planet semi-major axis ($a_p$) is much smaller than the warp location, their equation 4 can be simplified and the planet mass can be approximated by
\begin{equation}
    M_p = 0.8 M_{\rm Jup}  \left(\frac{\tau}{\rm 10\ Myr}\right)^{-1} \left(\frac{a_p}{\rm 10\ au}\right)^{-2} \left(\frac{a_w}{\rm 50\ au}\right)^{7/2} \left(\frac{M_{\star}}{1.8\ M_{\odot}}\right)^{1/2}.
\end{equation}
Figure~\ref{fig:warp} shows the planet mass as a function of semi-major axis to create a warp at 40 (blue) and 60~au (orange) after 10-17~Myr of secular interactions. Since the location of the warp is not well constrained, we use 40 and 60~au as a reasonable range consistent with the tentative warp reported by \cite{kasper2015}. Unfortunately, current limits for this system are poor and only rule out planets more massive than $\sim8$~$M_{\rm Jup}$ at 50-300~au projected separations \citep{wahhaj2013, meshkat2015}. Similarly, this system does not show a significant proper motion anomaly when comparing Hipparcos and Gaia eDR3 astrometry \citep{brandt2021, kervella2022}. Since this system is edge-on, these limits do not rule out the presence of a massive companion in the system at 1-20~au semi-major axes.

\begin{figure}[h!]
    \centering
    \includegraphics[width=0.5\textwidth]{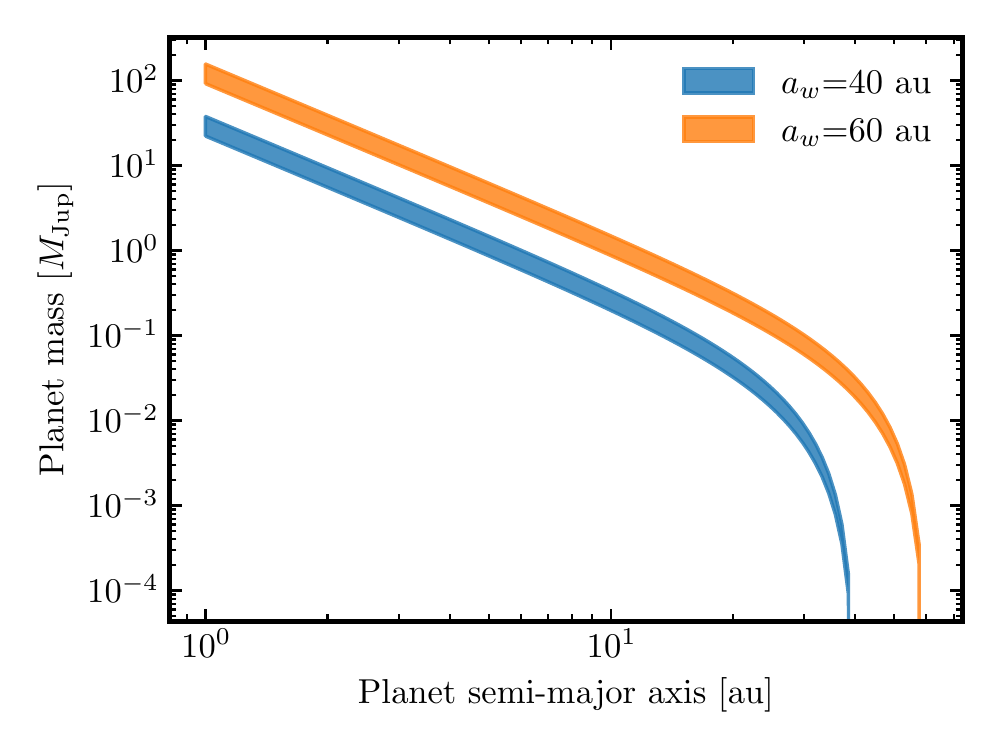}
    \caption{Planet mass and semi-major axis required to produce a warp at 40 (blue) or 60~au (orange). The width of the lines account for a secular evolution during 10-17~Myr.}
        \label{fig:warp}
\end{figure}

Finally, it is also interesting to compare the vertical distribution
of small dust. Although \cite{kasper2015} and \cite{esposito2020} did
not constrain the vertical distribution of small dust, the disk
appears to be flatter than in the ALMA observations that trace the
large dust. Small grains having a smaller inclination dispersion could
be a result of damping collisions \citep{pan2012} or even gas drag as
the dimensionless stopping time (Stokes number) of $\mu$m-sized grains
could be close to 1 (see \S\ref{sec:distribution_gas}). Forward
modelling of scattered light observations is needed to constrain the
vertical distribution of small dust and confirm this difference in
vertical distributions. It is also possible that the distribution of
small grains appears flatter due to the reduction methods to subtract
the stellar PSF, which can affect the observed morphologies.


\subsection{CO gas distribution}

\label{sec:distribution_gas}
The observations confirm the gas detection from \citet{Lieman2016},
and detect all four targeted carbon monoxide isotopologue transitions. Since gas is released from solid bodies (if secondary),  they should be roughly
co-located. We find that the CO gas emission is notoriously more compact, spanning between r$\sim1-10$~au out to 30~au, but with a peak radius at 10-20~au that is still consistent with the dust peak radius ($31_{-8}^{+10}$~au). The different distributions of the CO gas and dust is not unexpected. Models of gas-rich debris disks show that the
gas can viscously expand, reaching regions closer in and further out
unless the viscosity is very low \citep{kral2019,Marino2020}. However, these
models also show a sudden drop in the surface density of CO beyond
a critical outer radius where the carbon surface density drops to
levels in which it does not shield CO effectively from
photo-dissociation by interstellar UV. This destroys CO molecules at
large radii where column densities are low. 

In order to test if the gas and dust distributions can be reconciled in a secondary origin scenario, we use \textsc{exogas}\footnote{https://github.com/SebaMarino/exogas} \citep{Marino2020, marino2022} to model the radial evolution of gas released from the planetesimal belt. We consider a star with a mass of 1.8~M$_{\odot}$,  luminosity of 9~L$_{\odot}$, surrounded by a belt of planetesimals with a surface density that peaks at 22~au and with inner and outer FWHM's of 6 and 82~au\footnote{We modified \textsc{exogas}  to allow for a planetesimal belt with an asymmetric Gaussian distribution}. Note that these are the parameters that give the best fit to the continuum data and are slightly different from the medians presented in Table \ref{rtdusttable}. CO gas is input at a rate of $4\times10^{-3}\ $M$_{\oplus}/$Myr \citep[consistent with its fractional luminosity of $\sim 10^{-3}$ and a CO mass fraction of 10\% in planetesimals,][]{matra2017b}, and we evolve the gas for 10~Myr (a rough estimate of the period over which CO has been released after the protoplanetary disk dispersal) considering CO photodissociation, shielding by CO and CI, viscous spreading, and radial diffusion. We assume CI and CO are segregated with CI mainly present in a surface layer surrounding the CO gas creating optimal shielding \citep[analogous to assuming  negligible vertical diffusion][]{marino2022}. In addition, we update \textsc{exogas} such that the release rate of CO gas is also inversely proportional to the orbital period \citep[$\dot{\Sigma}_{\rm CO}^{+}\propto\Sigma_{\rm dust}^2 \Omega_{\rm K}$ with $\Omega_{\rm K}$ with the Keplerian frequency,][]{wyatt2007}. This enhances the CO release at smaller radii. Finally, we neglect the stellar UV 
in the CO photodissociation calculations. In reality, the stellar UV flux for this A-type star will be higher than the ISRF at the distances considered. However, we expect that the CO photodissociation at the gas disk inner edge will quickly form an optically thick CI layer in the radial direction that will shield the CO beyond that radius. \ah{It is important to note that selective photodissociation could reduce the abundance of $^{13}$CO  during
both the protoplanetary disk stage  \citep[while CO ices form, e.g. ][]{miotello2014}  and the debris disk stage \citep[while CO is released from solids; ][]{moor2019,cataldi2020}. Since the optical depth and mass are mainly constrained by the $^{13}$CO 
emission, a lower abundance would mean that the optical depth and mass of $^{12}$CO  are even higher than estimated.}

Figure \ref{fig:viscous} shows the evolution of the CO and CI gas for different viscosities (parametrized through $\alpha$). We find that CO can easily become shielded given the assumed CO released rate at the belt location. In order to fit the CO gas mass derived ($>10^{-2}~$M$_{\oplus}$ at 95\% confidence), $\alpha$ needs to be smaller or similar to 1$0^{-3}$. However, in order to explain the radial span of CO (10-30~au as highlighted by the vertical dashed lines) we find $\alpha\lesssim10^{-4}$. Note that while CO extends out to 60~au, its density drops exponentially beyond its peak near 20~au due to the CO release rate that decreases with radius (represented by the grey shaded area) and the CO lifetime that decreases with radius due to the lower surface density of CO (i.e. lower self-shielding). Therefore, we conclude that the compact nature of the CO emission is consistent with the observed distribution of mm dust in a secondary origin scenario.

\begin{figure}[h!]
    \centering
    \includegraphics[width=1.0\textwidth]{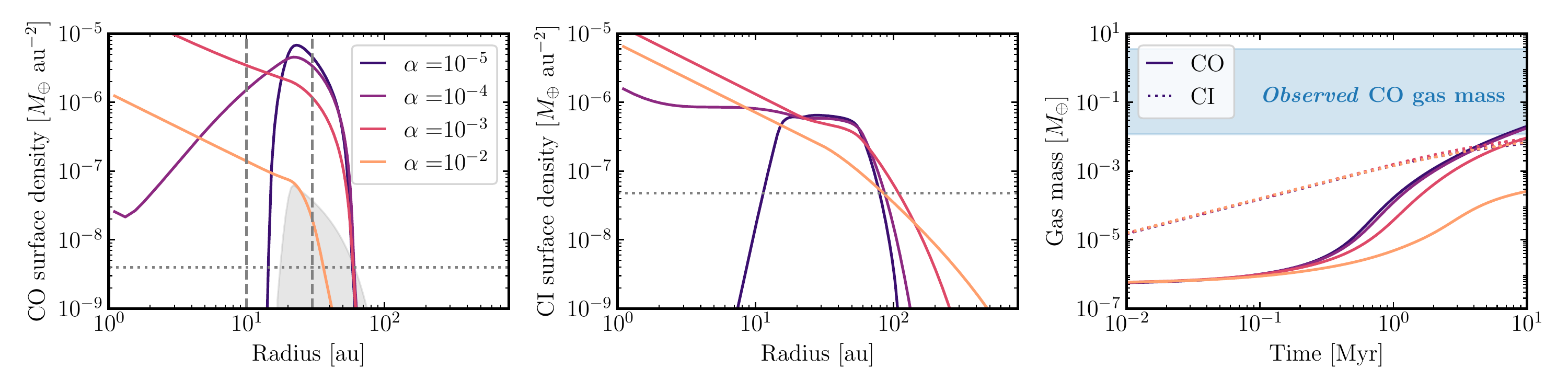}
    \caption{Simulated evolution of CO and CI undergoing viscous spreading and CO photodissociation using \textsc{exogas}. The surface density of CO and CI are shown in the left and middle panels for different viscosities after 10~Myr of evolution, while the right panel shows the temporal evolution of the total CO and CI mass. The vertical dashed lines in the left plot indicate the region where CO is significantly detected (10-30~au). The grey shaded region in the left panel represents the rate at which CO is released per unit area as a function of radius. The horizontal dotted lines in the left and middle panels represent the surface density above which CO starts to become self-shielded and shielded by CI. The blue shaded region in the right panel represents the CO mass derived from observations (95\% confidence level).}
        \label{fig:viscous}
\end{figure}

In addition to the effects considered, 
there could be other factors that could make the release of CO gas even more enhanced at smaller radii. The higher temperature of solids at smaller radii could increase the release rate of CO \citep[]{jewitt2017}, but also the release rate of CO$_2$, which can quickly photodissociate into CO+O and contribute to the CO gas \citep[]{lewis1983}. These two effects would make the CO distribution  appear even more compact as it would be heavily dominated by the innermost regions of the belt, perhaps matching even better the observations. Considering these effects is beyond the scope of this paper, but could be important when trying closely match the observations.

Finally, although we can fit the inner cavity in the CO gas distribution with a low viscosity, it is possible that the cavity exists due to a massive planet that is accreting most of the inflowing gas \citep{Marino2020, kral2020}. Such a planet could be the same that is responsible for the dust large vertical thickness and tentative warp (\S\ref{sec:vertical_dust}).

\subsection{Gas and dust interactions}

The morphology of the small dust detected by the scattered light
images could indirectly provide further information on the system's
total gas densities. If the gas densities are small, small dust is
created where the mm dust is and extends further out due to radiation
pressure. If the gas densities are high enough, small dust has Stoke\ah{s}
numbers close to 1 and can migrate out very quickly due to gas-drag
\citep{Takeuchi2001,Olofsson2022}. This effect is the same radial drift that
mm-sized dust experiences in protoplanetaty (Class~II) disks. In this
case, dust migrates out because small dust is even more sub-Keplerian
than gas due to the effect of stellar radiation pressure , which is not the case of protoplanetary disks where the stellar radiation is completely blocked below the disk's surface. If gas densities are
even higher, the Stokes number would be $\ll$1, in which case the small
dust would be coupled and will follow the gas as in a
protoplanetary disk. 

In order to quantify if this could be the case in HD~110058, we perform a similar analysis to the one in \S5.6 in \cite{Marino2020} to estimate if radial migration could be important for the dynamics of small grains. Figure~\ref{fig:Tr} shows the migration timescale from 20 to 30~au of the small grains at the blow-out limit ($\sim2\mu$m) relative to their collisional lifetime as a function of the gas surface density. The red section of the line shows the densities that are ruled-out by our observations (at 95\% confidence). If CO dominates the surface density over carbon and oxygen (as found in our model in \S\ref{sec:distribution_gas}), we expect that the gas density will be in the range $2\times10^{-4}-7\times10^{-2}$~g~cm$^{-2}$ and thus the radial migration timescale relative to the collisional timescale will be smaller than one (green section). If this is the case, we would expect a pile up of small dust near the outer edge of the gas disk between 20-30~au before the density drops significantly. In this green section the smallest grains will have Stokes numbers close to 1 and thus will tend to settle towards the midplane as recently shown by \cite{Olofsson2022}. Therefore, the scale height of small grains could be significantly smaller than the scale height of large grains. If the gas was primordial and thus dominated by other than CO (e.g. H or H$_{2}$), then the gas densities would be expected to be a factor $\geq10^3$ higher than derived for CO, leading to densities above $\sim0.1$~g~cm$^{-2}$ (blue section). In this case the small dust grains are well coupled to the gas, decreasing its migration rate and settling towards the midplane (as in protoplanetary disks). In this high gas density regime gas drag could reduce the relative velocities of small grains and slow down the removal of unbound sub-blowout grains \citep{Lecavelier1998}. Both effects could increase significantly the overall abundance of small grains and the disk fractional luminosity. \ah{The scattered light observations presented in \citet{esposito2020} suggest that HD110058 may be reproduced by a disk with a moderately small aspect ratio, constant with radius, down to 16 au from the star. Unfortunately, \citet{esposito2020}  did not include HD~110058 in their disk morphology modelling of the scattered light images and we cannot confidently say anything conclusive from the images they provide. New SPHERE data (Stasevic et al. private communication) will provide a detailed analysis of the scattered light emission.}



\begin{figure}[h!]
    \centering
    \includegraphics[width=0.5\textwidth]{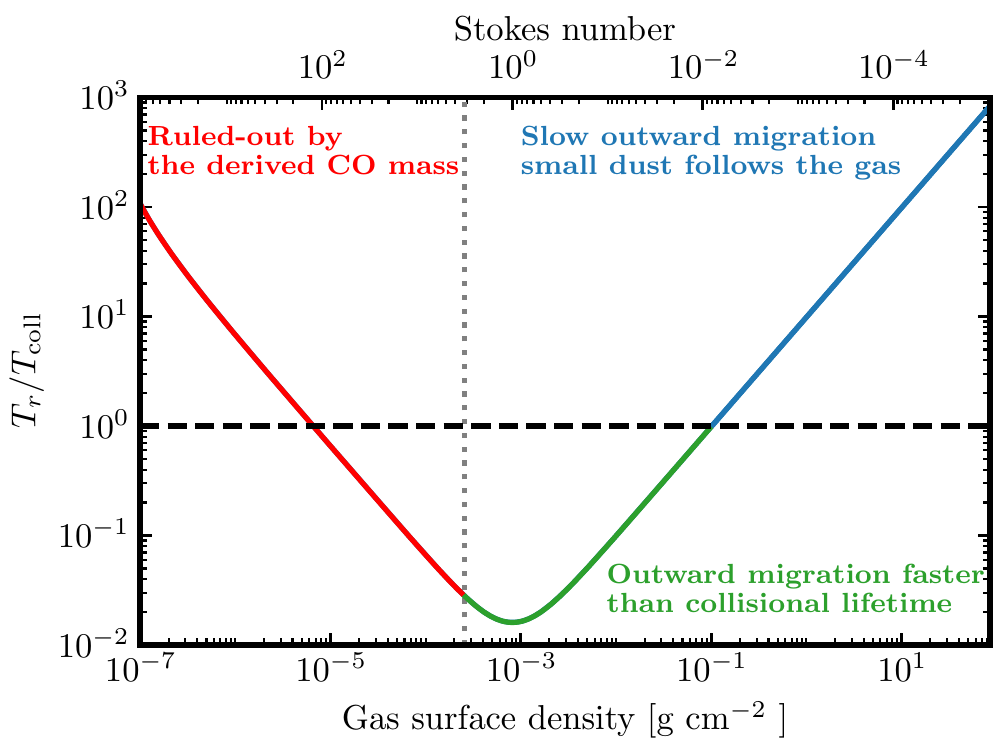}
    \caption{Radial migration timescale relative to the collisional lifetime of small grains as a function of gas surface density (solid line) following \cite{Marino2020} and using the derived parameters for HD~110058. The horizontal dashed lines represents a migration timescale equal to the collisional timescale. The vertical dotted line shows the minimum gas surface density derived from the minimum CO gas mass and radial distribution (95\% confidence). The red section of the solid line represents the regime ruled out by our observations. The green section represents the region consistent with the CO observations and where the outward migration would be faster than the collisional timescale. The blue section represents the region where the gas density is high enough (and dominated by other gas species than CO  such that the migration is very slow as small dust becomes well coupled to the gas.}
        \label{fig:Tr}
\end{figure}

\section{Conclusion}\label{conclusion}

This work presents ALMA Band 6 and 7 observations of the HD110058 gas-rich debris disk. The observations detect the disk in continuum, $^{12}$CO and $^{13}$CO. The disk is among the most compact debris disk around early-type stars observed by ALMA so far. We used radiative transfer model to characterize the distributions of dust and gas, and discuss the results in the context of evolutionary models of debris disks. Our main findings are:

\begin{itemize}

\item The dust disk is compact with a peak radius of 31~au, but with a very smooth outer edge that leads to a FWHM of 49~au and a large fractional width of 1.2. The disk's inner edge of roughly 18~au (smaller than 23~au with 99.7\% confidence), is the smallest debris disk's inner edge inferred with ALMA so far. 

\item  We found the dust disk's inclination is i=78$^{+9}_{-12}$~$^\circ$. If we impose that $i>80^{\circ}$ to be consistent with scattered light observations and the gas modelling, we find that the disk must be vertically resolved with h=0.12-0.28. This would imply an inclination dispersion for the solids of 11-23$^{\circ}$, consistent with the  Kuiper belt's classical hot population and scattered disk. This is also consistent with the smooth outer edge, which could be due to high eccentricities.

\item The total dust and gas masses derived using radiative transfer models are $0.080_{-0.003}^{+0.002}$~\mearth\   and $0.069^{+3.56}_{-0.057}$~\mearth, respectively. We also find the best fit stellar mass is 1.84$_{-0.18}^{+0.16}$ M$_{\odot}$, suggesting the star is a late A- dwarf (A6/7V) instead of A0V (consistent with recent optical measurements of the stellar temperature). 

\item The CO gas distribution is more compact than the dust ($\le$10 to 30~au), but with a peak radius consistent with the dust's peak radius. The distributions of dust and gas can be explained with models of radial gas evolution released by collisions in the planetesimal belt, thus favoring the secondary origin scenario.

\end{itemize}
 
Deeper observations to trace the innermost extension of the gas, or of the micron-sized  dust which indirectly traces the gas if the densities are sufficiently high, are required to further constrain the gas densities and the overall history of the system.

\software{Common Astronomy Software Applications \citep{McMullin2007}, {\sc RADMC-3D} \citep{Dullemond2012}, GALARIO, \citep{tazzari2017}, EMCEE \citep{foreman2013}, Astropy \citep{2013A&A...558A..33A}, {\sc pdspy} \citet{Sheehan2019}, \textsc{exogas}, \textsc{disc2radmc} \citep{Marino2020, marino2022}, Starlink Software Collection \citep{currie2014}.}

\section*{Acknowledgments}

This paper makes use of the following ALMA data: ADS/JAO.ALMA\#2018.1.00500.S. ALMA is
a partnership of ESO (representing its member states), NSF (USA) and
NINS (Japan), together with NRC (Canada) and NSC and ASIAA (Taiwan),
in cooperation with the Republic of Chile. The Joint ALMA Observatory
is operated by ESO, AUI/NRAO and NAOJ. The National Radio Astronomy
Observatory is a facility of the National Science Foundation operated
under cooperative agreement by Associated Universities, Inc. S. M. is supported by a Junior Research Fellowship from Jesus College, University of Cambridge. S.P. acknowledges support from FONDECYT grant 1191934 and funding from ANID -- Millennium Science Initiative Program -- Center Code NCN2021\_080. L.M. acknowledges funding from the European Union’s Horizon 2020 research and innovation programme under the Marie Sklodowska-Curie grant agreement No. 101031685.

{}

\appendix

\section{Position-Velocity Diagrams}\label{pvdiag}

\begin{figure}
    \centering
    \includegraphics[scale=0.4]{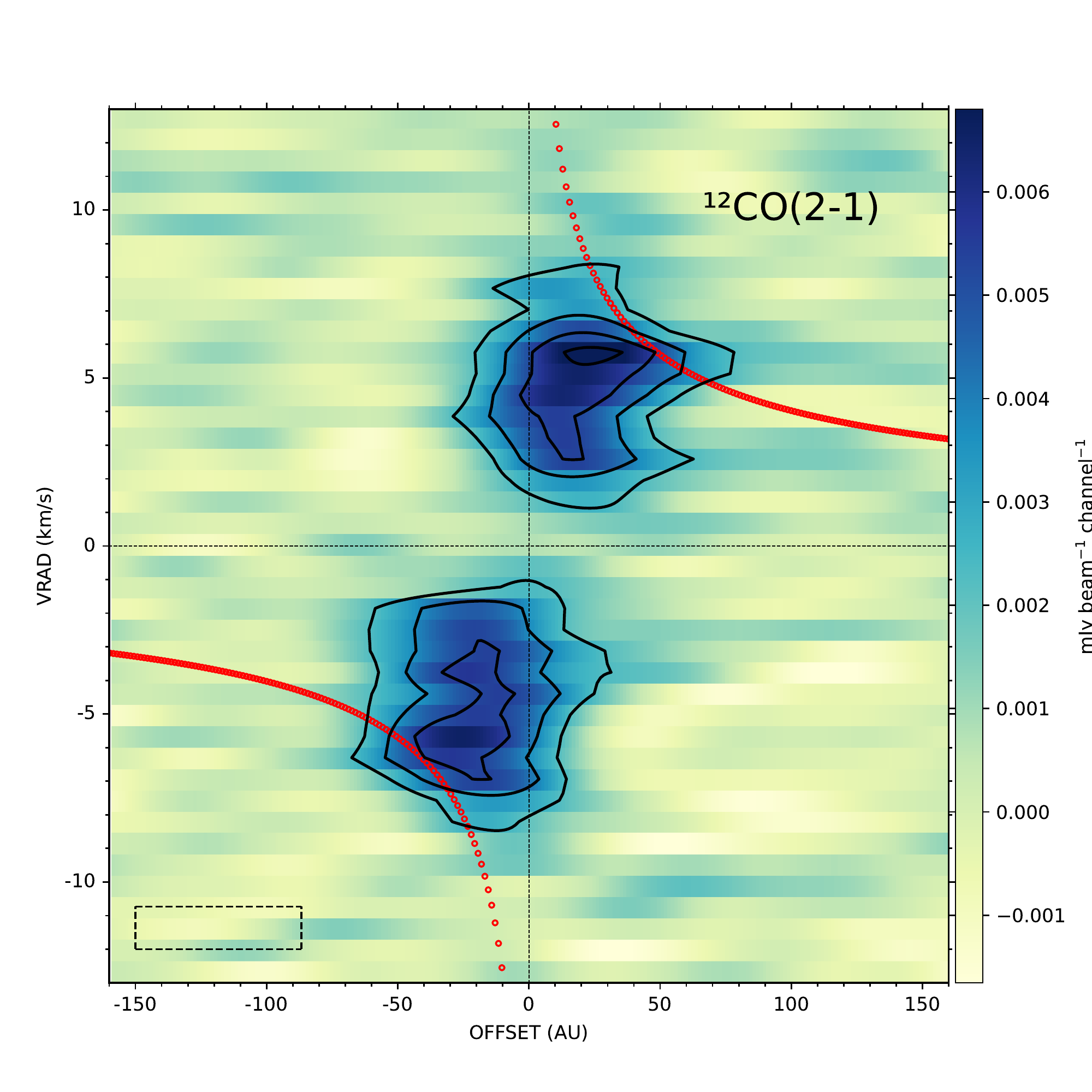}
    \includegraphics[scale=0.4]{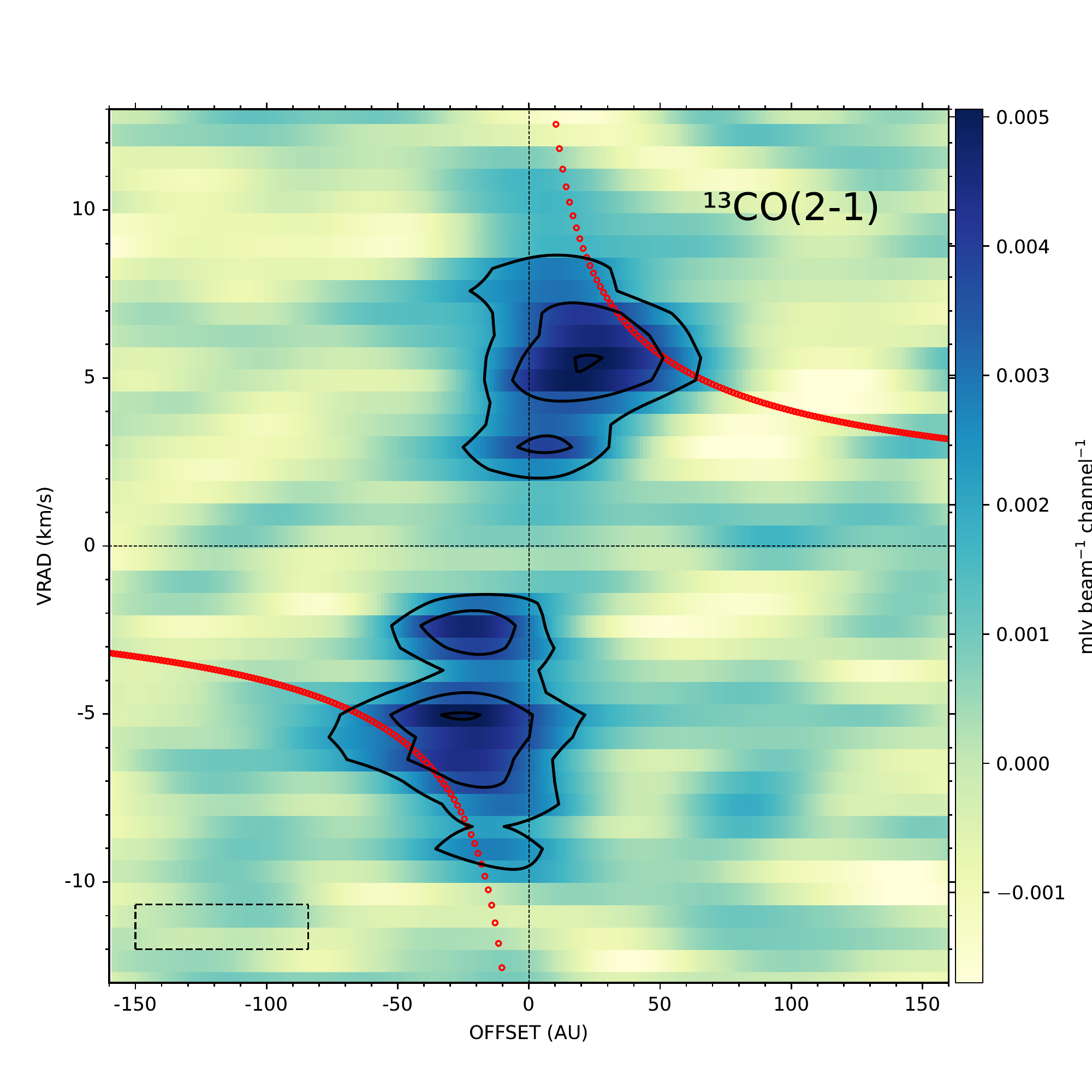}
    \includegraphics[scale=0.4]{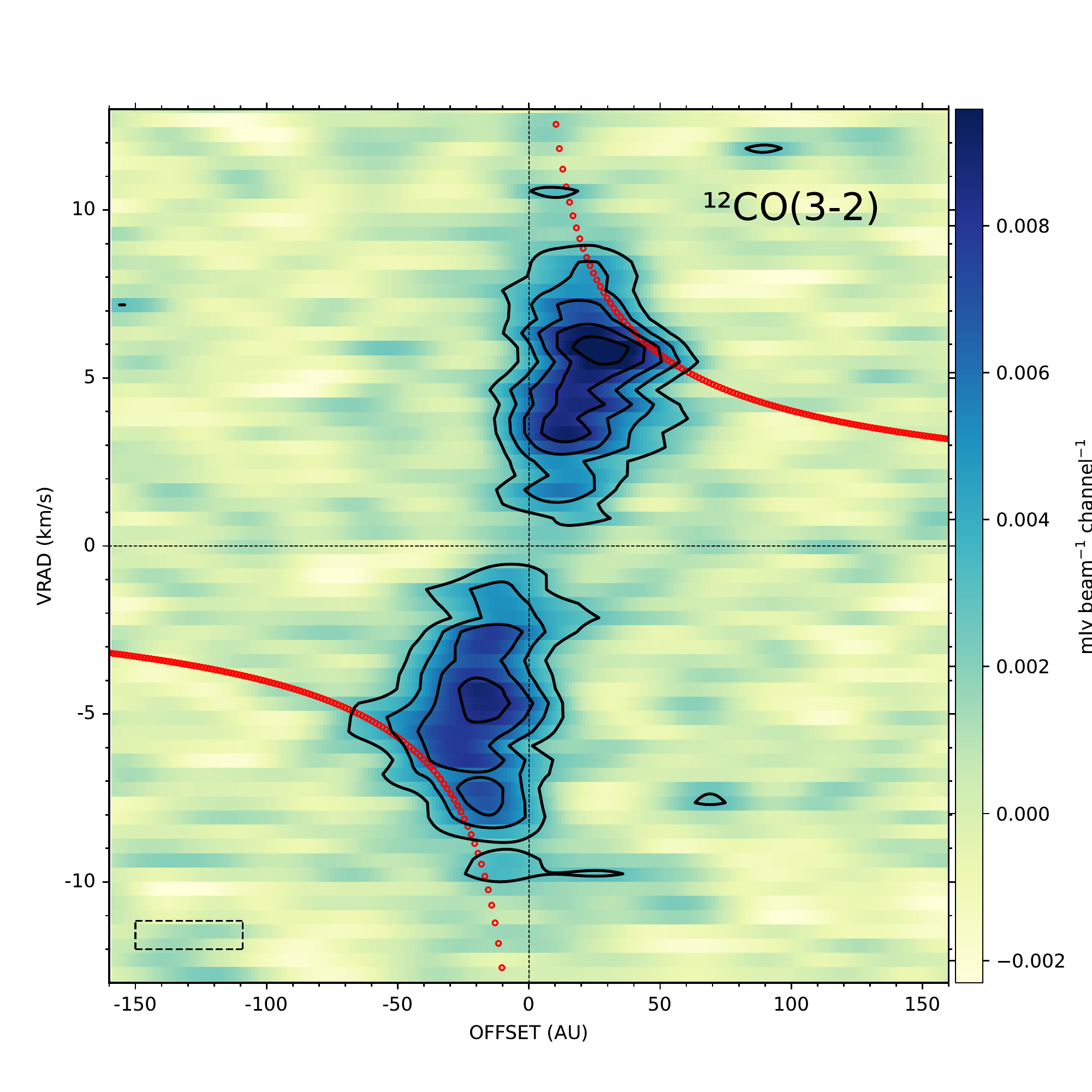}
    \includegraphics[scale=0.4]{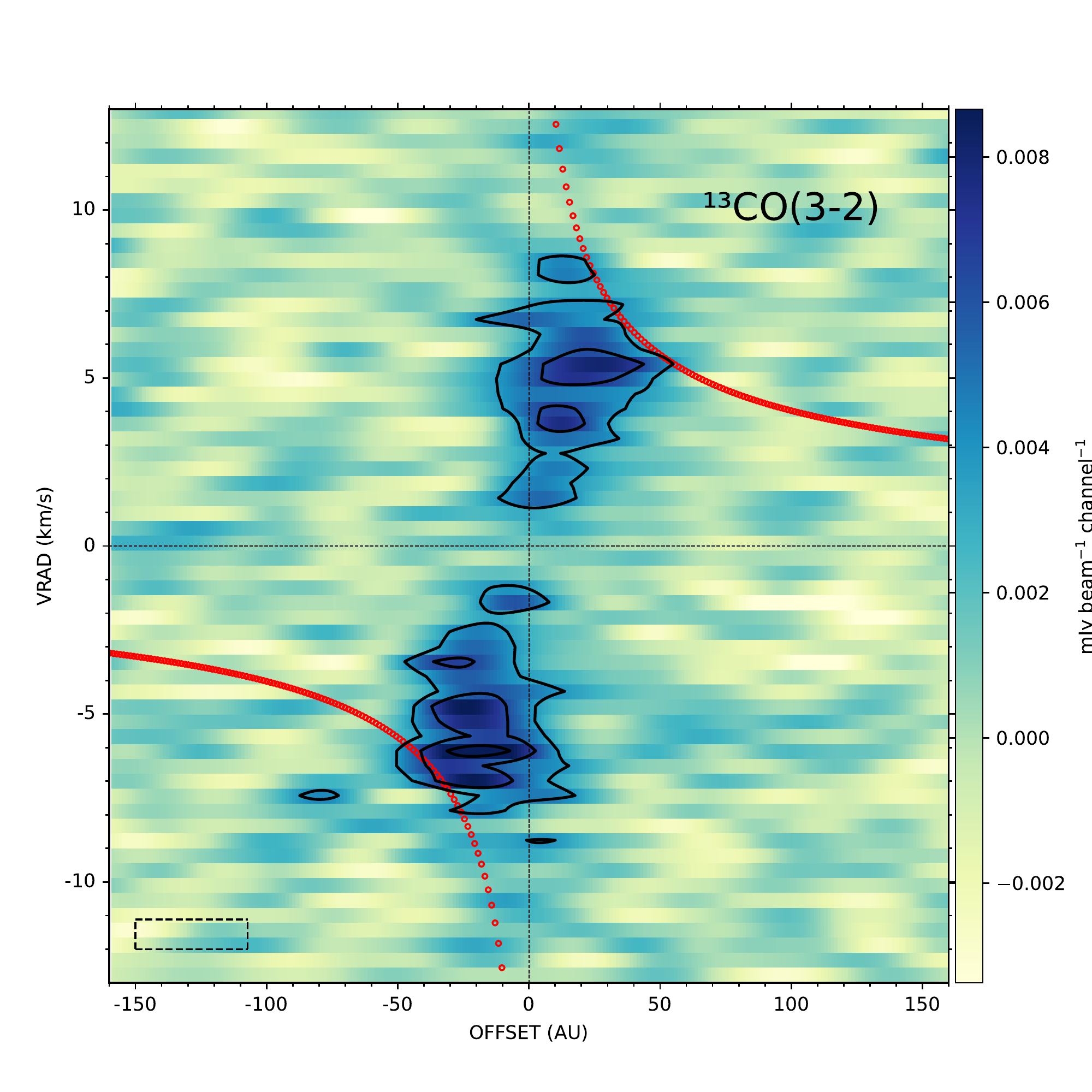}
    
    \caption{ Top: Position-Velocity diagram of $^{12}$CO (2$-$1)
      (left) and $^{13}$CO (2$-$1) (right). The red dots are the
      maximum keplerian velocities, and the black contours levels at 3, 6, 9, 12
      times the RMS \ah{(0.79 and 0.73~mJy~beam$^{-1}$, respectively)}.  Bottom: Same for $^{12}$CO (3$-$2)
       and $^{13}$CO (3$-$2), left and right panel respectively. \ah{The RMS for  $^{12}$CO (3$-$2)
       and $^{13}$CO (3$-$2) are 1.11 and 1.53~mJy~beam$^{-1}$. }
    }
    \label{pvdiag_fig}
\end{figure}

We produced Position-velocity (PV) diagrams using Astropy and
SpectralCube libraries in Python. Position-velocity slices were
extracted by integrating $\sim$25~au (0.19\arcsec) above and below the
midplane disk (in the direction perpendicular to the disk's major
axis). Figure~\ref{pvdiag_fig} shows the PV diagrams for the different
molecular transitions.

Similar to \cite{matra2017a} we compute the ratio of the PV diagrams of
the different molecules/transitions (the ratios between the (3-2) and
(2-1) transitions of each molecule, and the ratios between the two
molecules in same transition).

For this we produced new cubes with TCLEAN, in which the angular and
spectral resolution of Band 7 data are degraded in order to match the
resolution of the Band 6 datasets. The PV diagrams of each molecule
were combined in order to compute the different ratios in spaxels with
signal higher than 4$\sigma$. Figure~\ref{intraband_ratios} shows the
ratio between the PV diagrams of $^{12}$CO(3$-$2) and
$^{12}$CO(2$-$1), and $^{13}$CO (3$-$2) and $^{13}$CO(2$-$1).
Figure~\ref{interband_ratios} shows the ratio between the PV diagrams
of $^{12}$CO and $^{13}$CO for each transition. Figure~\ref{pv_tau}
shows the resulting optical depths. 

\begin{figure}
    \centering
    \includegraphics[scale=0.3]{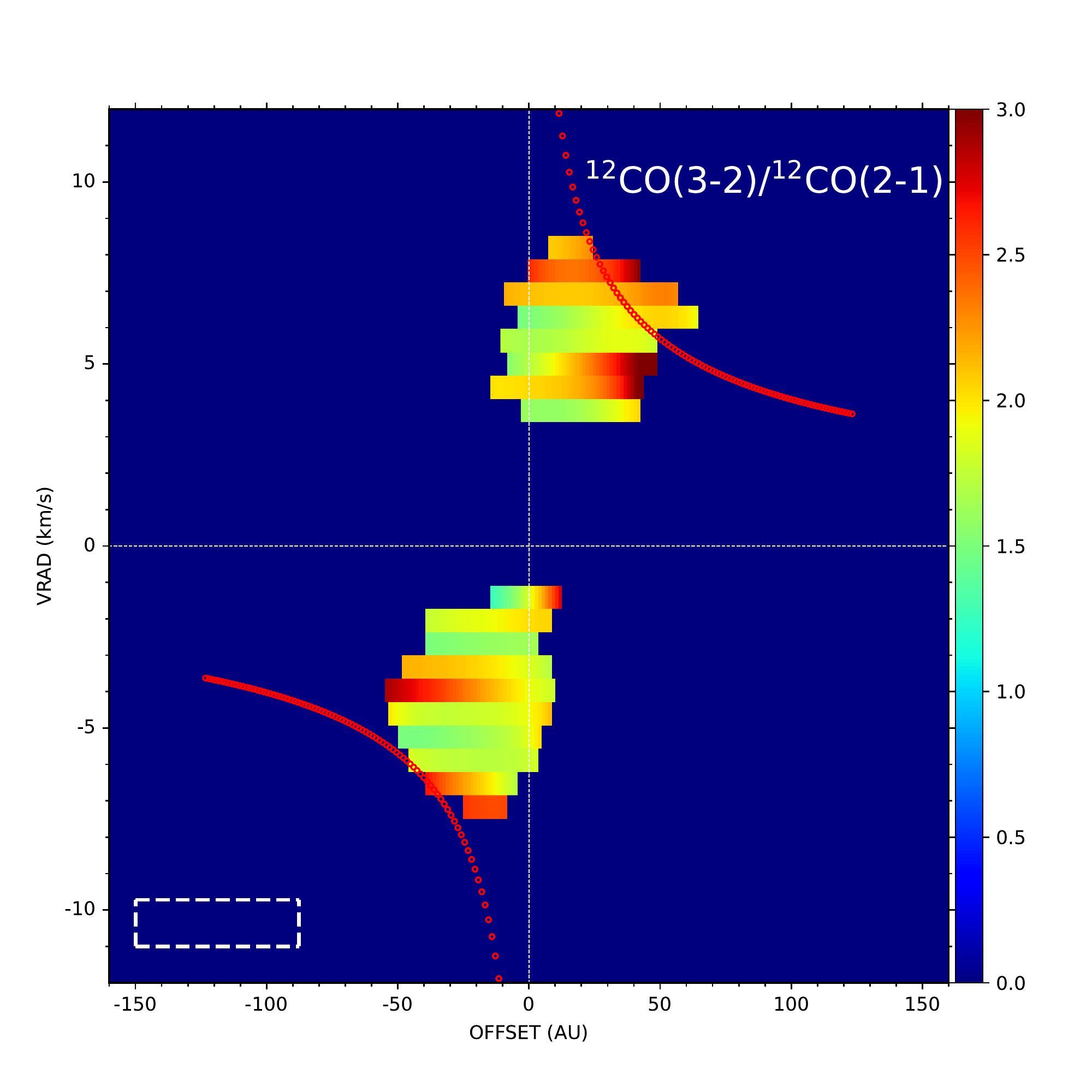}
    \includegraphics[scale=0.3]{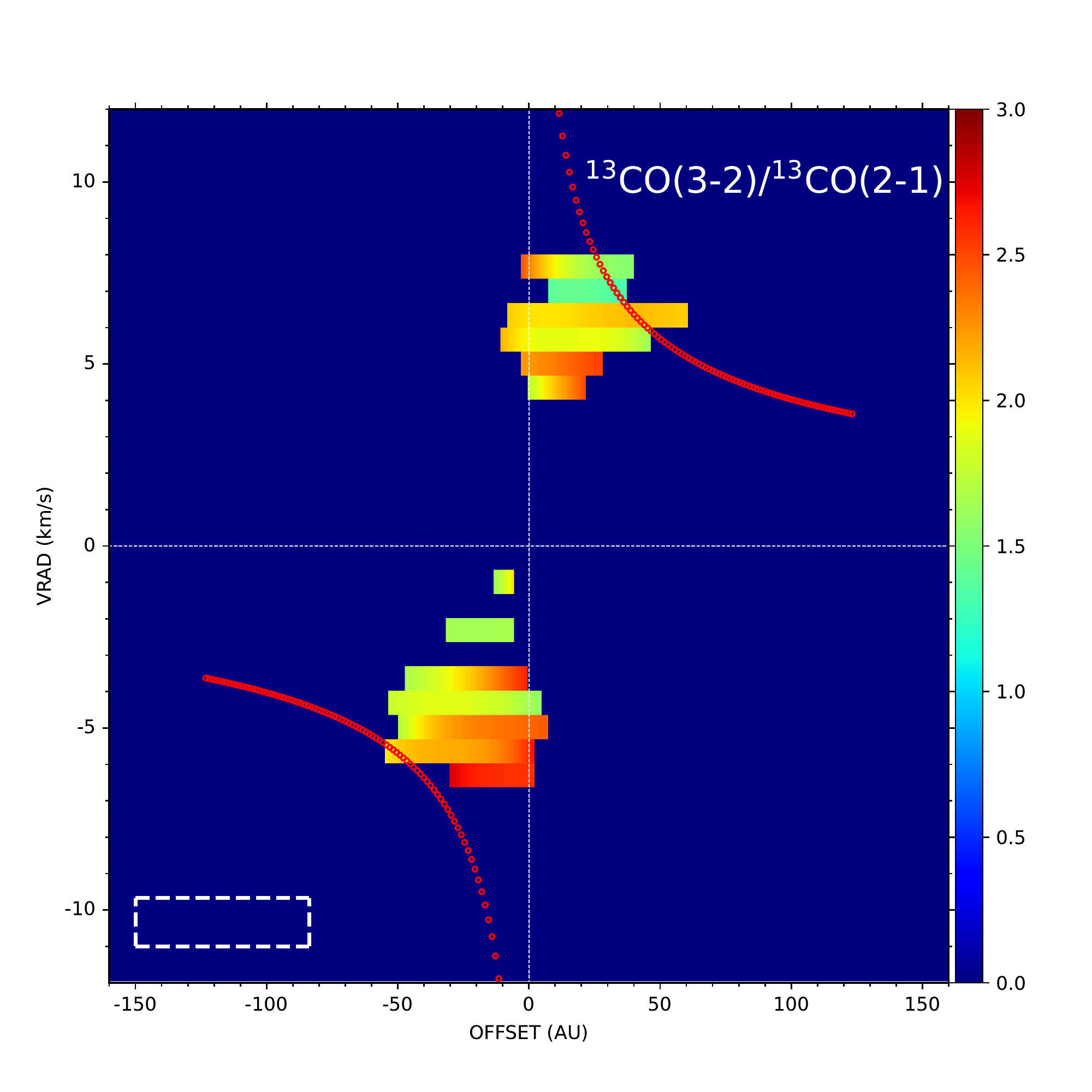}

    \caption{Ratio of the $^{12}$CO(3$-$2) and $^{12}$CO(2$-$1) P-V
      diagrams (left). Ratio of the $^{13}$CO (3$-$2) and $^{13}$CO
      (2$-$1) P-V diagrams (right). The ratios are computed in spaxels where both P-V diagrams
      have signal higher than 4$\sigma$.  }
    \label{intraband_ratios}
\end{figure}

\begin{figure}
    \centering                  
    \includegraphics[scale=0.3]{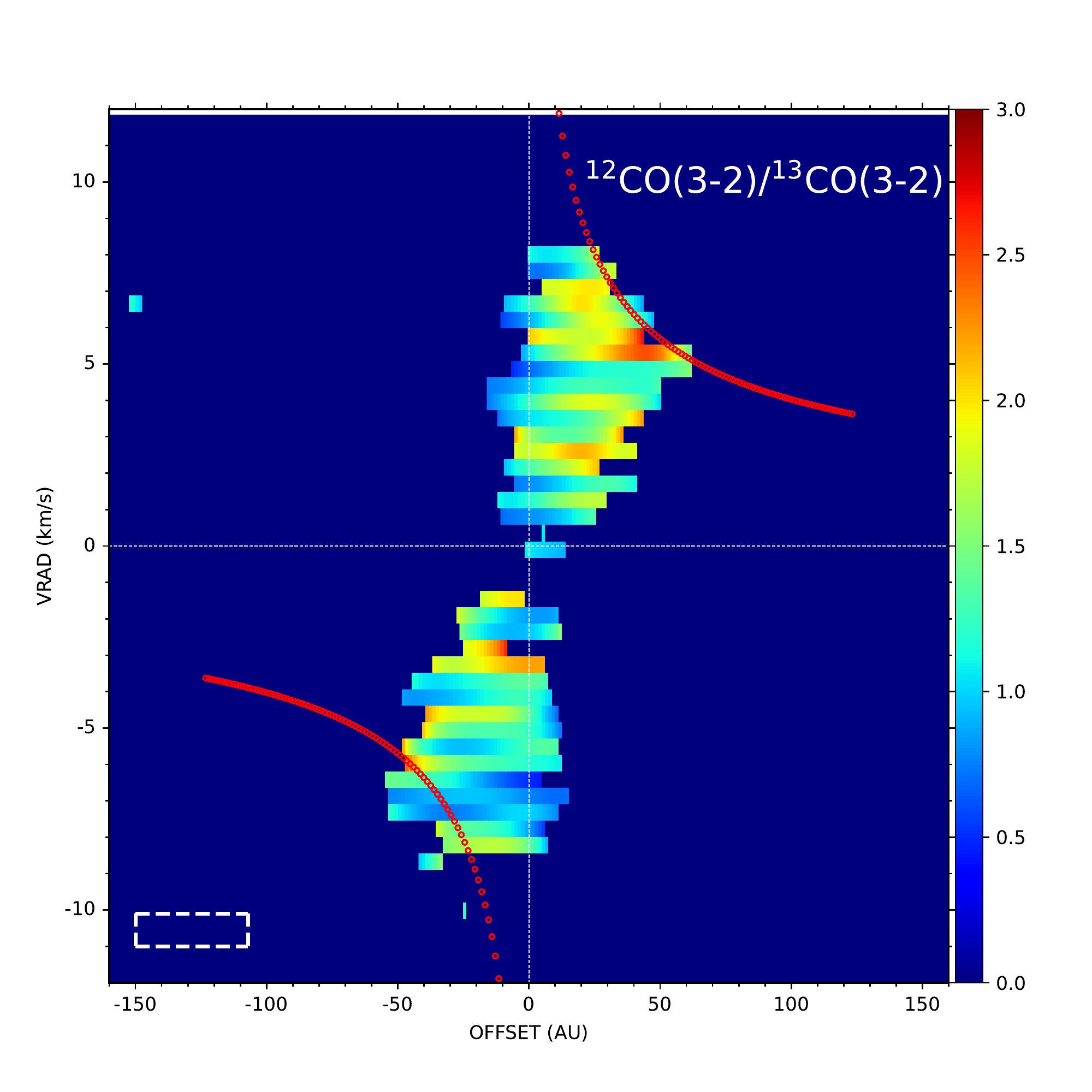}
    \includegraphics[scale=0.3]{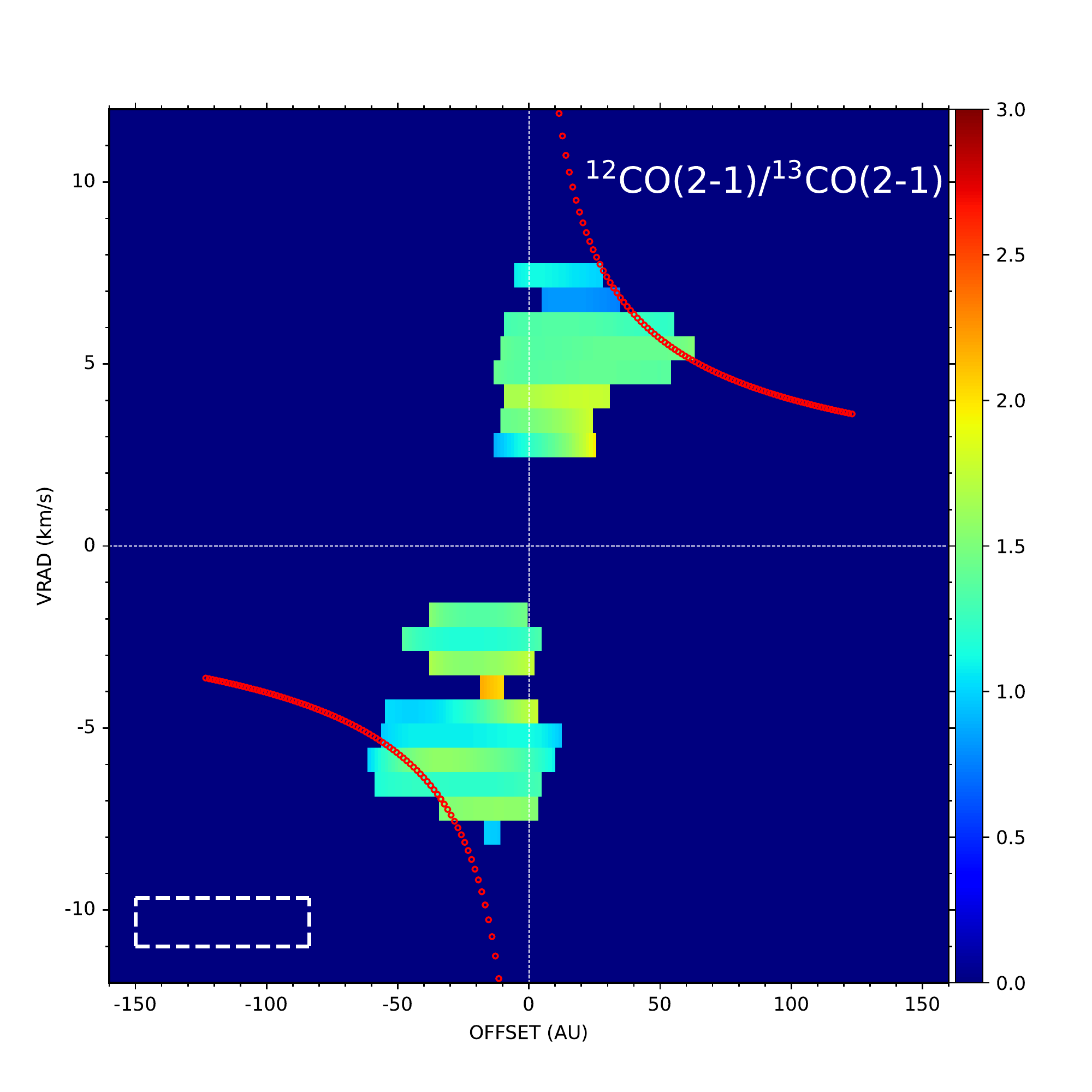}

    \caption{Position-Velocity diagram of $^{12}$CO(3$-$2)/$^{13}$CO(3$-$2)
      (left) and $^{12}$CO(2$-$1)/$^{13}$CO(2$-$1) (right).  The ratios are computed in spaxels where both P-V diagrams
      have signal higher than 4$\sigma$. 
    }
    \label{interband_ratios}
\end{figure}

\newpage

\begin{figure}
    \centering
    \includegraphics[scale=0.4]{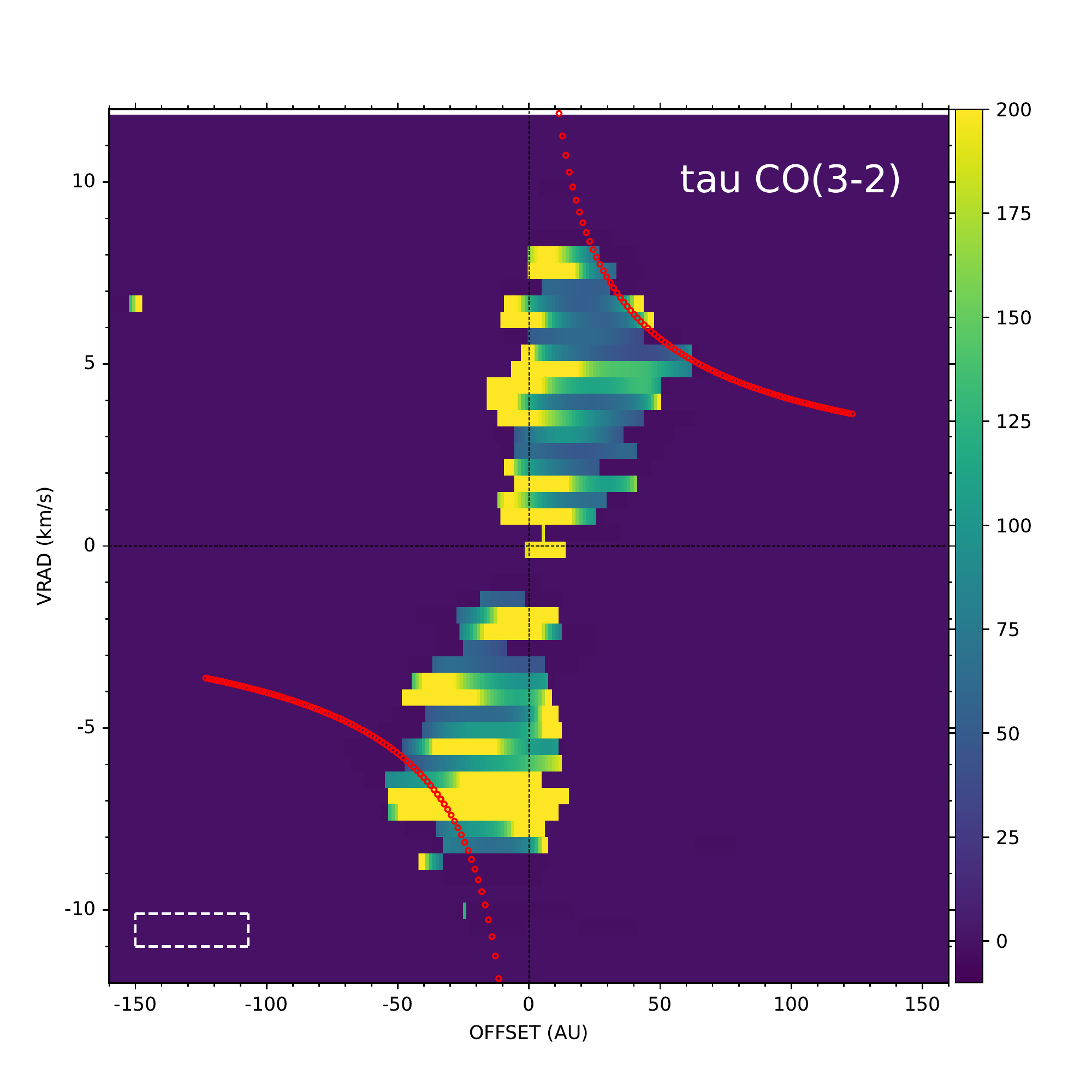}
    \includegraphics[scale=0.4]{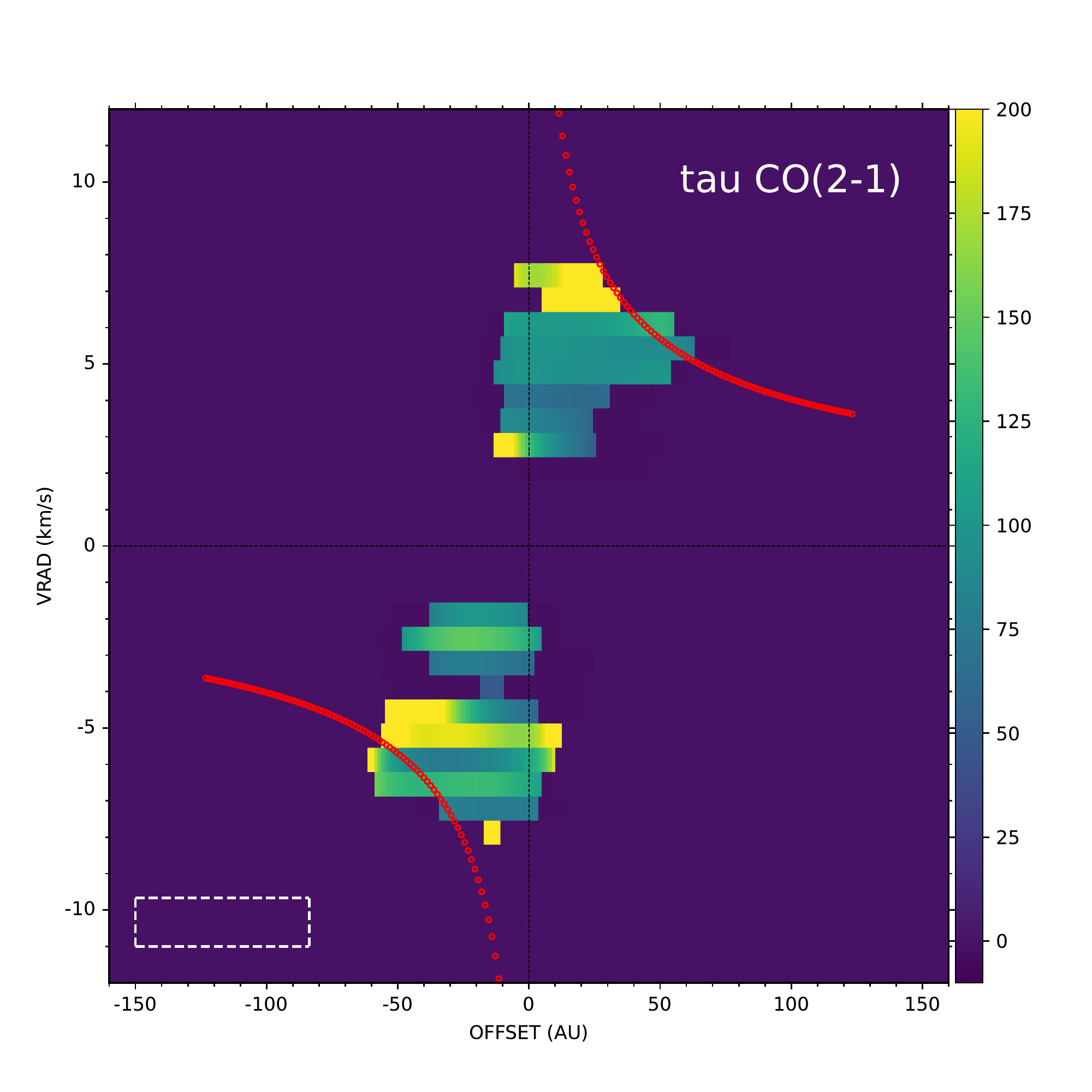}

    \caption{P-V diagram of $^{12}$CO optical depth from (3$-$2) (left) and CO(2$-$1) (right). The   $^{12}$CO optical depth $\tau$ was obtained by solving $R\times(1-e^{-\tau/76}) = 1-e^{-\tau}$, where R is the  $^{12}$CO/ $^{13}$CO line ratio.      
    }
    \label{pv_tau}
\end{figure}

%
%
%
%
%
%

\section{Channel maps and gas models. }\label{uztaumodelgas}

\begin{figure}
\begin{center}
\epsscale{0.5}

\includegraphics[angle=0,scale=0.7]{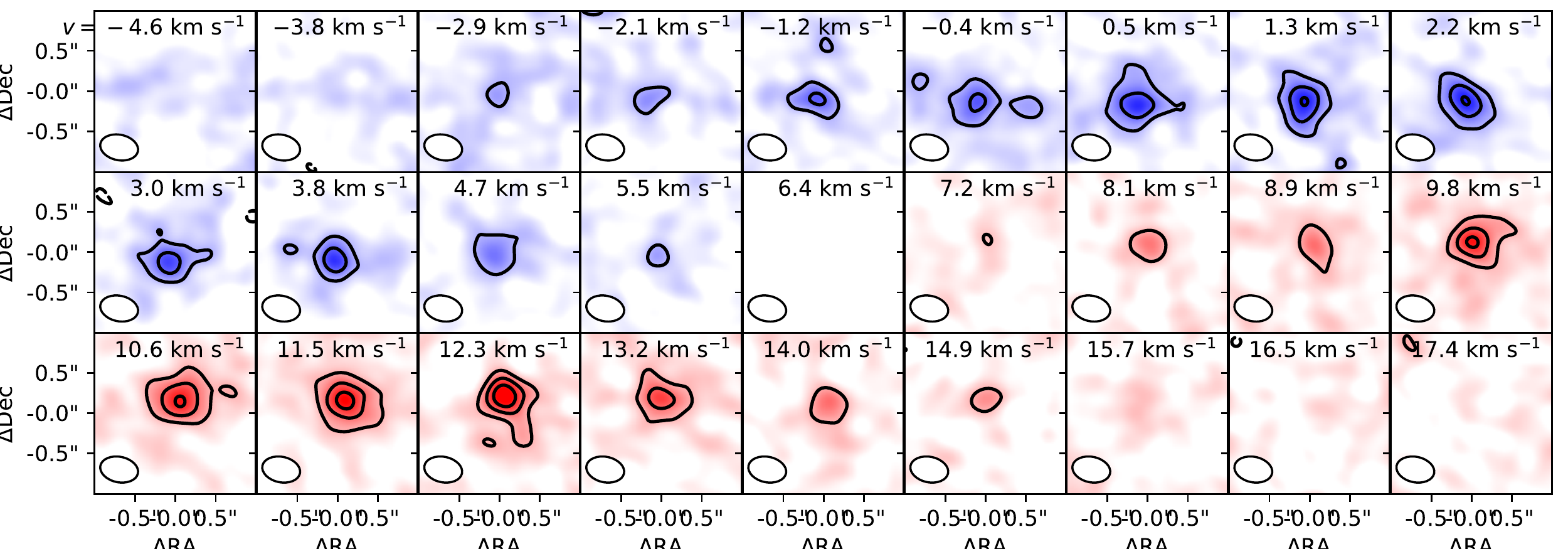}
\includegraphics[angle=0,scale=0.7]{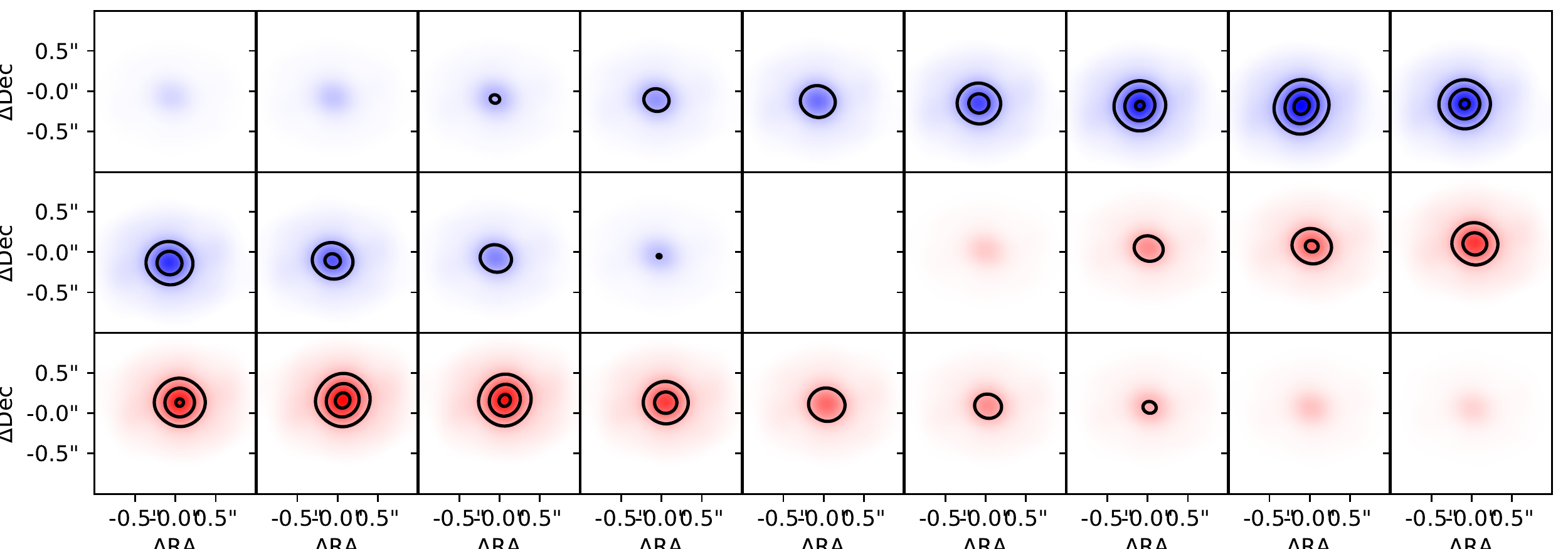}
\includegraphics[angle=0,scale=0.7]{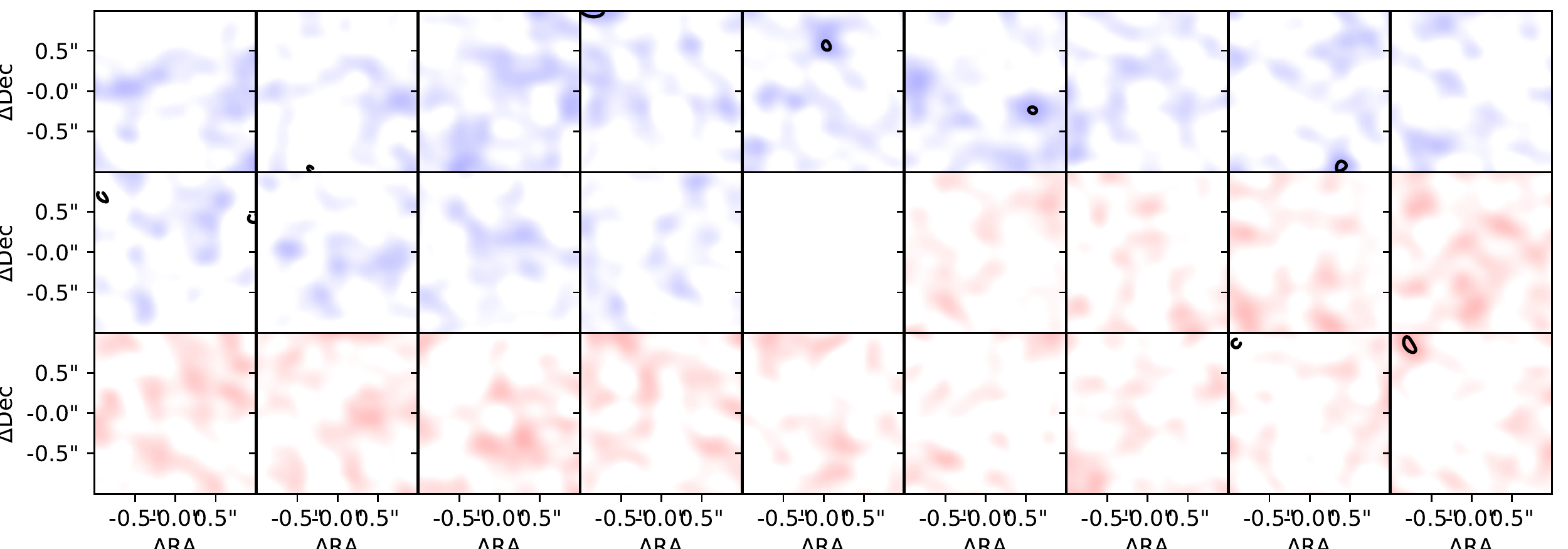}

\caption{ $^{12}$CO(3-2) channel maps for HD~110058, best-fit disk model (middle) and residuals (bottom). The model image is generated by using GALARIO to Fourier transform the best-fit model image onto the same baselines as the ALMA visibility data, and then imaged using a CLEAN implementation built into pdspy. The residual image is similarly made by subtracting the best-fit model visibilities from the data and then imaging. We show solid contours starting at \ah{$3\sigma$}, with increments of $3\sigma$ \ah{(the RMS per channel is 1.11~mJy~beam$^{-1}$)}. Similarly, dashed contours show emission starting at \ah{$-3\sigma$} and continuing in increments of $-3\sigma$.}
\label{fig:mcmc12co32}
\end{center}
\end{figure}

\begin{figure}
\begin{center}
\epsscale{0.5}

\includegraphics[angle=0,scale=0.7]{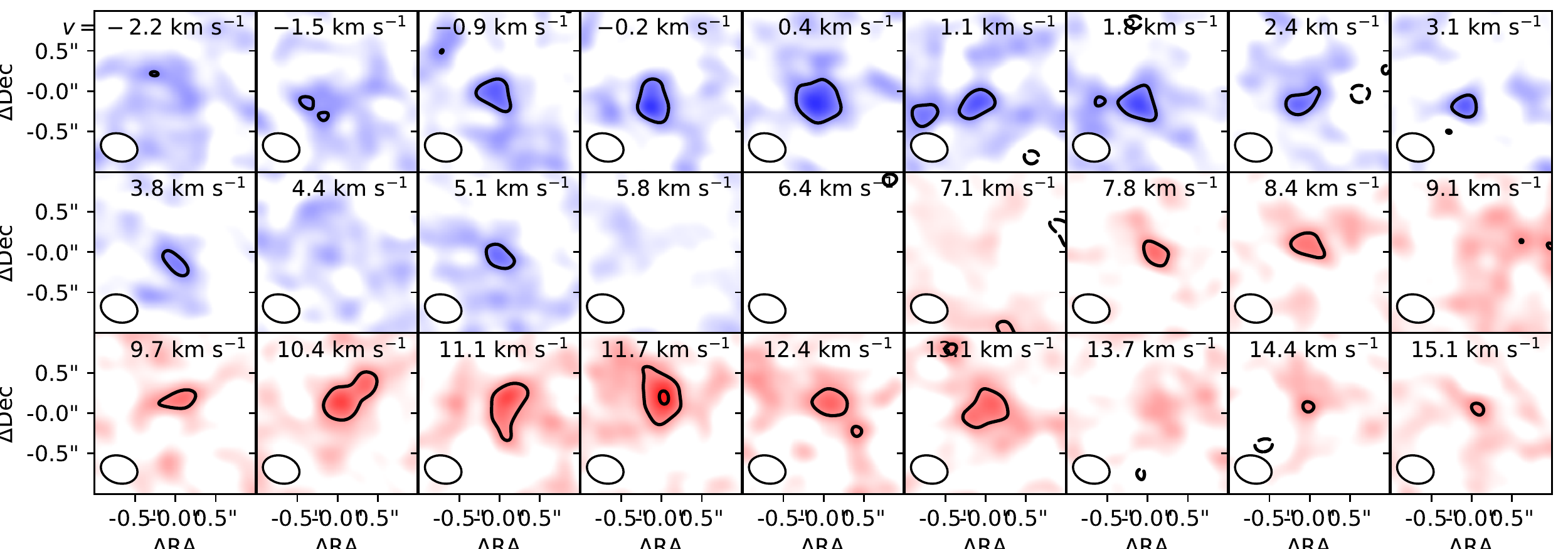}
\includegraphics[angle=0,scale=0.7]{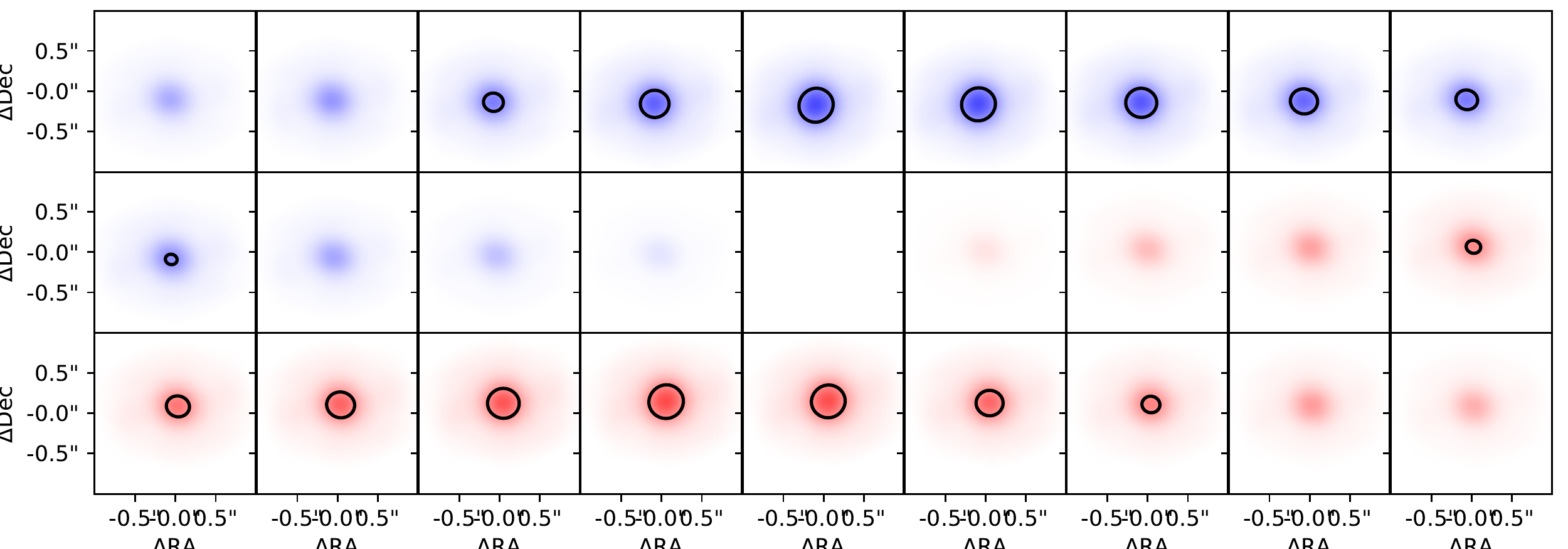}
\includegraphics[angle=0,scale=0.7]{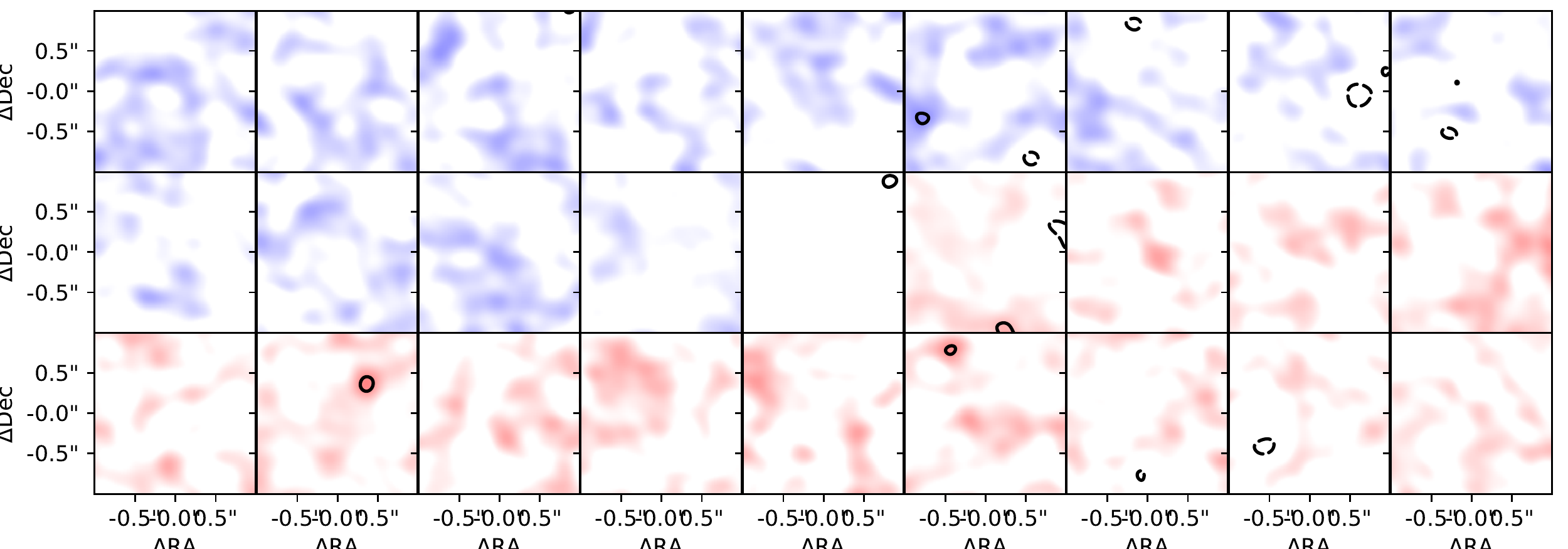}

\caption{$^{13}$CO(3-2) channel maps for HD~110058, best-fit disk model (middle) and residuals (bottom). Contour levels are the same as for Figure~\ref{fig:mcmc12co32}. \ah{The RMS per channel is 1.53~mJy~beam$^{-1}$.}}  
\label{fig:mcmc13co32}
\end{center}
\end{figure}

\begin{figure}
\begin{center}
\epsscale{0.5}

\includegraphics[angle=0,scale=0.7]{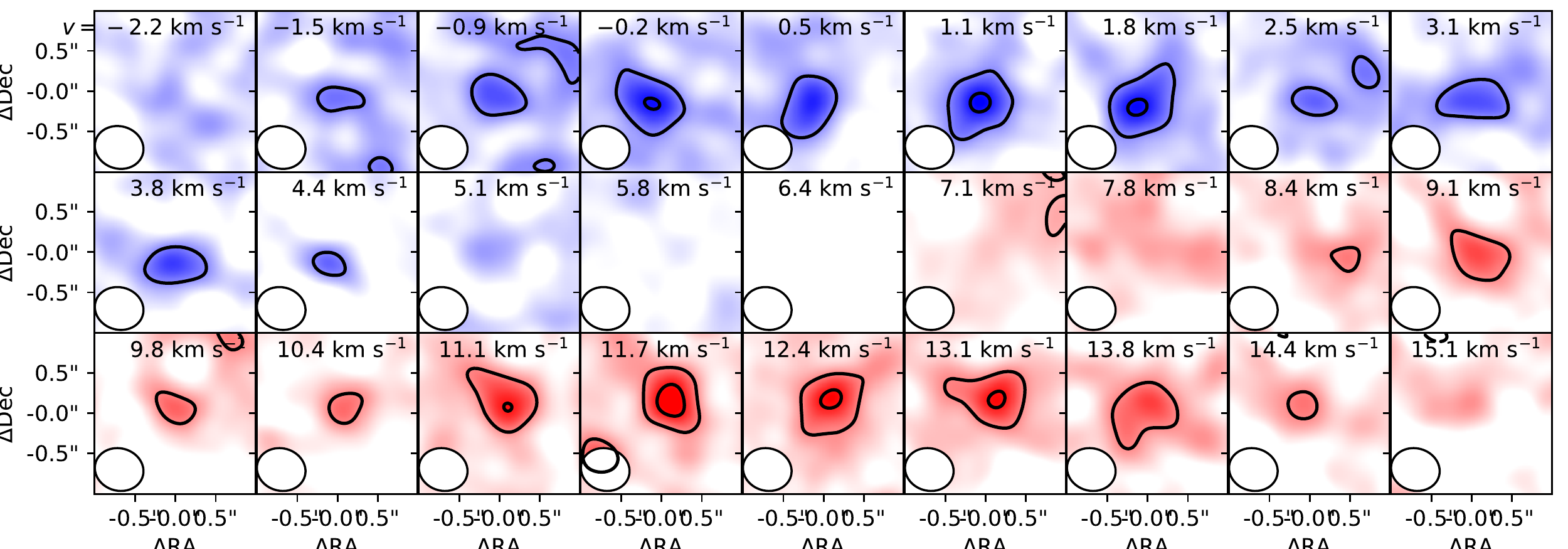}
\includegraphics[angle=0,scale=0.7]{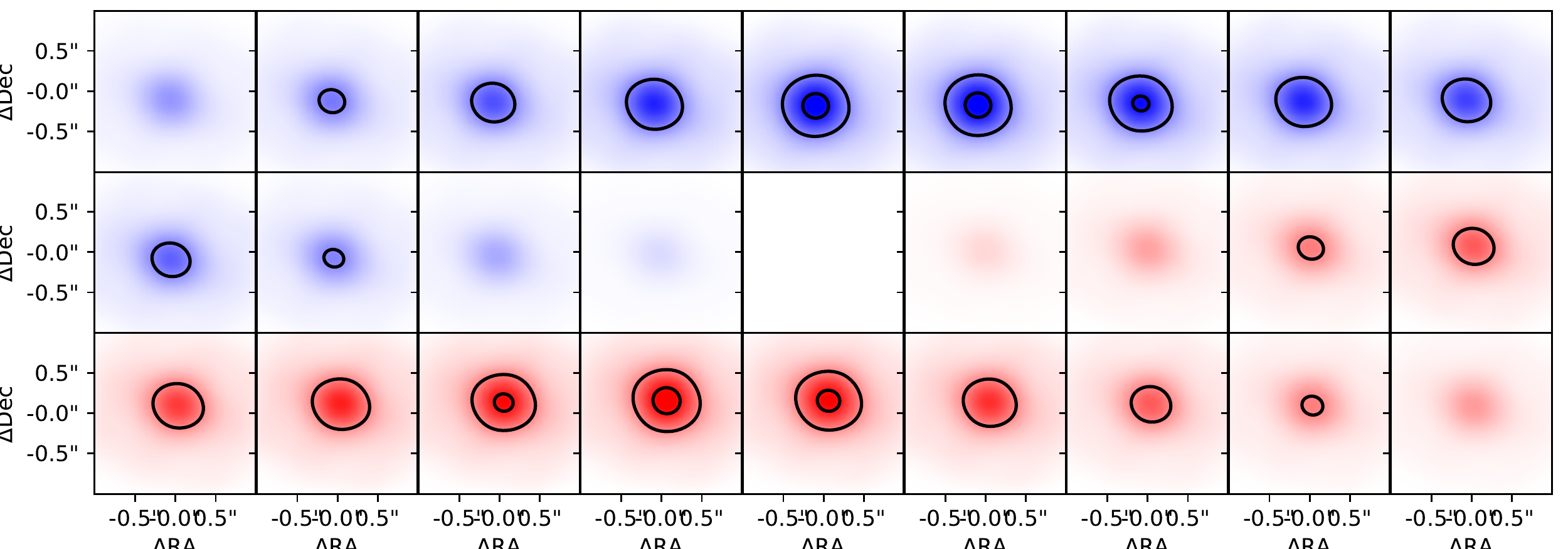}
\includegraphics[angle=0,scale=0.7]{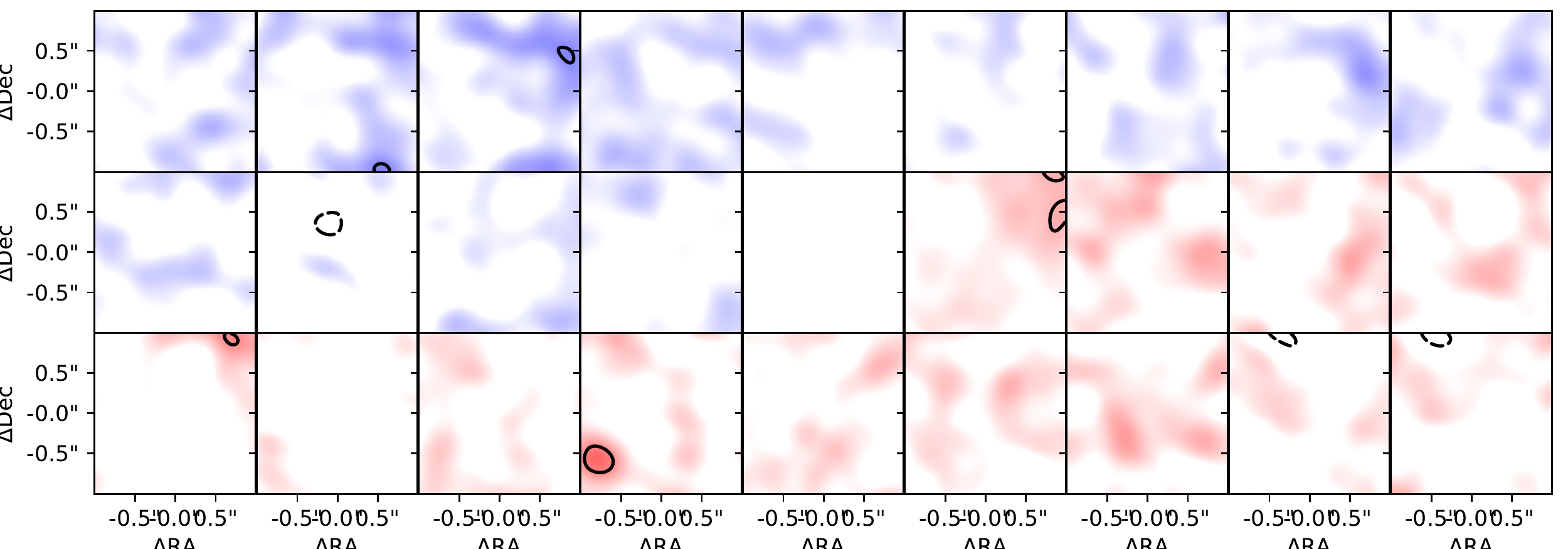}

\caption{$^{13}$CO(2-1) channel maps for HD~110058, best-fit disk model (middle) and residuals (bottom). Contour levels are the same as for Figure~\ref{fig:mcmc12co32}. \ah{The RMS per channel is 0.73~mJy~beam$^{-1}$.}}  
\label{fig:mcmc13co21}
\end{center}
\end{figure}

%
%
%

%
%

\end{document}